\documentclass[aps,reprint,pre,groupedaddress,superscriptaddress,showpacs,longbibliography]{revtex4-1}

\usepackage{natbib}
\usepackage[latin1]{inputenc}
\usepackage{graphicx,amsmath,amssymb,stmaryrd,bm,dcolumn,textcomp,braket}
\usepackage{mathrsfs}
\usepackage{rotating,array,color,float}
\usepackage{xcolor,siunitx}             
\usepackage{hyperref}
\usepackage[inline]{showlabels}
\usepackage{cancel}
\hypersetup{colorlinks,linkcolor={blue!80!black},citecolor={blue!80!black},urlcolor={blue!80!black}}

\newcommand{\q}[2]{\bm{q}{}{}^{#1}_{#2}}

\newcommand\Tstrut{\rule{0pt}{2.5ex}}         % = `top' strut
\begin{document}

\title{Quantifying Acoustophoretic Separation of Microparticle Populations by Mean-and-Covariance Dynamics for Gaussians in Mixture Models}

\author{Fabio Garofalo}
\email[corresponding author:]{fabio.garofalo@bme.lth.se}
\affiliation{Department of Biomedical Engineering, Lund University,
Ole R\"omers V\"ag 3 S-22363, Lund, Sweden}

%\author{Umberto Picchini}
%\email{umberto.picchini@bme.lth.se}
%\affiliation{Department of Mathematics, Lund University,
%Ole R\"omers V\"ag 3 S-22363, Lund, Sweden}

\date{\today}

\begin{abstract}
A method for the quantification of acoustophoretic separation and dispersion for microparticle populations featuring continuously distributed physical parameters is presented.
The derivation of the method starts by (i)~considering the equation of motion for a particle ensemble in the coordinate+parameter space, (ii)~performing moment analysis on the transport equation for the probability density function (PDF), and (iii)~expanding up to the first-order the drift (and the diffusion coefficient) around the mean of the PDF.
Following these steps, a system of ordinary differential equations for the evolution of the mean and the covariance in the coordinate+parameter space is derived.
%The dynamics of the spatial marginal for particle populations during acoustophoretic separation can be thus approximated by the dynamics of gaussians in a mixture model in the coordinate+parameter space.
%Since the dynamics of the PDF is the mean and the covariance of the PDF is known, it is possible to 
These differential equations enable for the approximation of the acoustophoretic separation dynamics of particle ensembles by using a gaussian mixture for which the mean and the covariance of each gaussian evolve according to the mean-and-covariance dynamics.
The approximation property of this method is shown by comparison with direct numerical simulations of particle ensembles in the cases of prototypical models of acoustophoretic and free-flow acoustophoretic separations for which the particle populations are distributed according to the radius.
Furthermore, the indicators for quantifying free-flow acoustophoretic separation performance are introduced, and a method for the inference of particle-histogram parameters is illustrated.
\end{abstract}

\maketitle

\section{Introduction}
Acoustofluidics is a microfluidic technology that using acoustic waves is able to perform separation of microbeads and viable manipulation of cells~\cite{Burguillos_2013, Wiklund_2012}.
Indeed, by exploiting the interaction between acoustic pressure waves and a carrier-fluid suspension of microbeads/cells at microscale~\cite{Bruus_2012_Acoustofluidics_7, Settnes_2012, Karlsen_2015}, acoustophoresis~\cite{Bruus_2011} is able to trap~\cite{Evander_2012}, wash~\cite{Augustsson_2012}, concentrate~\cite{Nordin_2012}, align~\cite{Manneberg_2008} and separate the suspended microparticles~\cite{Augustsson_2012_Anal_Chem, Ding_2014, Petersson_2007}.
The ability to separate microparticles is based on the different particle properties, such as compressibility and density. Specifically for cells, the different physical properties are associated with biological differentiation, type-uniformity and pathological conditions~\cite{Titushkin_2006, Cross_2007, Dao_2003, Remmerbach_2009, Constantino_2011}.

The microparticle physical parameters that appear in the acoustophoretic force expression~\cite{Bruus_2012_Acoustofluidics_7, Settnes_2012, Karlsen_2015} are not well-represented by unique values, e.g. single values for the radius, the compressibility and the density, but they occur as distributions for the microparticle populations.
Therefore, a model for the quantification of acoustophoresis must incorporate a mechanism that, taking into account for the statistics of the samples, allows to predict a continuous differentiation in the microparticle population trajectories and thus in the separation performance.
However, the present models of acoustophoretic trajectories rely on statisticsless descriptions that do not quantify the impact of the continuously distributed particle parameters on the separation performance~\cite{Muller_2012, Muller_2013, Garofalo_2014}.
Furthermore, since the acoustophoresis outcomes are directly related to the distribution of the physical properties, it is of interest to establish if (i)~assuming the knowledge of the device features by performing hydrodynamic and acoustic calibration, and (ii)~measuring the separation performance is possible to determine the distribution of the physical parameters for the particle population.

A possible and straightforward solution to overcome the drawbacks of the present models is to evolve particle ensembles that are normally distributed in both parameter and space~\cite{simon2017particle}.
However, the limitation of this kind of techniques becomes apparent when the parameter and/or the spatial distributions are not gaussians, and even more in parameter-estimation procedures which, being based on multiple calculations, must be extremely cheap in terms of the computational cost associated with a single calculation, i.e. the evolution of a single gaussian.

A more convenient method that (i)~evolves the mean and the covariance of a normally distributed ensemble~\cite{Garofalo_2014}, and (ii)~approximates the particle distribution by using a mixture model with gaussian kernels is proposed in this paper.
For that, the proposed method can be addressed as ``mean-and-covariance dynamics for gaussians in mixture models'', or briefly MCDGM.

A method for the evolution of the mean and the covariance of particle ensembles can be traced back to the stochastic linearization methods, that are widely used in mechanics~\cite{Elishakoff_2012, Socha_2007}, and recently for the quantification of dispersion in acoustophoretic models~\cite{Garofalo_2014}.
Stochastic linearization methods can include higher-order moments, but then closure assumptions are needed and the reconstructed PDF can violate the positivity assumption.
The only difference is that in modeling acoustophoresis of microparticle populations, the thermal fluctuation, i.e. Brownian motion, can be neglected as this becomes relevant only for nanoparticles.
For completeness, in this paper the derivation of the mean-and-covariance dynamics retains the diffusion term, that is dropped when the method is applied to acoustophoresis of microparticle population.
The mean-and-covariance dynamics can be also framed within the moment analysis techniques~\cite{Brenner_1993}, that, together with the PDF reconstruction, have been used in the context of quantifying the dispersion in microfluidic devices, such as in Deterministic Lateral Displacement separators~\cite{Cerbelli_2013, Cerbelli_2015}.
Since the MCDGM method can be used to approximate the actual particle distribution at the outlet section of the device, it also provides the indicators necessary to quantify the acoustophoretic separation performance.

In order to illustrate the application of the MCDGM method for the quantification of acoustophoretic separation of microparticle populations, this manuscript is organized as follow.
Section~\ref{sec:theory} (A)~reviews the derivation of the mean-and-covariance dynamics by (i)~introducing the equation of motion for a particle ensemble in the state space, (ii)~introducing the associated transport equation, and (iii)~performing moment analysis with linearization of the drift and the diffusion around the mean of the PDF.
Section~\ref{sec:theory} (B) (i)~translates the mean-and-covariance dynamics from the state space to the spatial+parameter space by providing the explicit expressions for the evolution of the spatial average and the spatial/mixed-covariance of a single gaussian and (ii)~introduces the gaussian mixture approximation for the parameter marginal and for the reconstruction of the spatial marginal.
Section~\ref{sec:examples} specializes the MCDGM method to the study of acoustophoretic separation by showing the comparisons with particle ensemble simulations for (A)~a prototypical model of acoustophoretic separation, (B)~the buffer-dependent separation of RBC and WBC similar to that presented in~\cite{urbansky2017rapid}, and (C)~the 3D simulations for free-flow acoustophoresis in a rectangular microchannel. Finally (D)~the application of the method in the estimation of particle size histogram is illustrated.

\section{Theory}\label{sec:theory}
\subsection{Mean-and-Covariance Dynamics}
Let us consider the nonlinear stochastic differential equation in the It\^o sense \cite{Risken_1996, Frank_2010}
\begin{subequations}\label{eq:sde}
\begin{align}
\label{eq:sdedyn} \mathrm{d}\bm{Q}(t)&=\,\bm{f}(\bm{Q},t)\,\mathrm{d}t+\sqrt{2}\,\bm{\sigma}(\bm{Q},t)\cdot
\mathrm{d}\bm{W}(t)\,,\\
\label{eq:sdeic} \bm{Q}(t_0^{})&=\,\bm{Q}_0\,,
\end{align}
\end{subequations}
where \mbox{$t,t_0\in\mathbb{R}$} with \mbox{$t>t_0$}, \mbox{$\bm{Q}_0,\bm{Q}(t)\in\mathbb{R}^d\equiv\mathbb{Q}$} are the realizations of the random process in the $d$-dimensional state-space, \mbox{$\bm{f}(\bm{q},t):\mathbb{R}^d\times \mathbb{R}^+\rightarrow \mathbb{R}^d$} is the drift,
and \mbox{$\bm{\sigma}(\q{}{} ,t):\mathbb{R}^d\times \mathbb{R}^+\rightarrow \mathbb{R}^{d\times d}$} is the standard deviation matrix.
The latter trasforms the differential of the multivariate Wiener process \mbox{$\bm{W}(t):\mathbb{R}^+\rightarrow\mathbb{R}^d$} defined by
\begin{subequations}\label{eq:wprocdef}
\begin{align}
\label{eq:wprocexp} \mathrm{E}[\bm{W}(\tau)]&=\,\bm{\bar{W}}=\,\bm{0}\,,\\
\label{eq:wproccov} \mathrm{cov}[\bm{W}(\tau)]&=\,\tau\,\bm{I}\,,
\end{align}
\end{subequations}
into the displacement $\mathrm{d}\bm{Q}$ for the states $\bm{Q}$.
In equation~\eqref{eq:wprocdef}, $\bm{I}$ is the unit tensor, \mbox{$\tau\in\mathbb{R}^+$} is a time-translation, \mbox{$\mathrm{E}[\bm{X}]$} and \mbox{$\mathrm{cov}[\bm{X},\bm{Y}]=\mathrm{E}[(\bm{X}-\bm{\bar X})(\bm{Y}-\bm{\bar Y})^T]$} are the expected value and the cross-covariance, respectively. The covariance can be written as \mbox{$\mathrm{cov}[\bm{X}]=\mathrm{cov}[\bm{X},\bm{X}]$}.
Equation~(\ref{eq:sdeic}) represents the initial condition for the realizations of Eq.~\eqref{eq:sdedyn} in terms of the realizations $\bm{Q}_0$ that is distributed according to a probability density function \mbox{$\rho(\q{}{0},t_0)$}.

Equation~(\ref{eq:sde}) corresponds to the (forward) Fokker-Planck equation for the probability density $\rho(\q{}{},t\,|\q{}{0}\,t_0)$~\cite{Risken_1996}
\begin{equation}\label{eq:fpfweq}
\partial^{}_t \rho(\q{}{},t\,|\,\q{}{0},t^{}_0)=\mathscr{L}^{}_{\mathrm{FP}}(\q{}{},t)\,\rho(\q{}{},t\,|\,\q{}{0},t_0)\,,
\end{equation}
conditioned for \mbox{$t=t_0$} by
\begin{equation}\label{eq:fpfwic}
\rho(\q{}{},t_0\,|\,\q{}{0},t_0)=\int_{\mathbb{Q}}\delta(\q{}{}-\q{}{0})\,\rho(\q{}{0},t_0)\,\mathrm{d}\q{}{0}\,.
\end{equation}
In equation \eqref{eq:fpfweq} the Fokker-Planck forward operator (assuming Einstein notation)
\begin{equation}\label{eq:fpfwop}
\mathscr{L}^{}_{\mathrm{FP}}(\q{}{},t)g=
-\partial^{}_{h}\left[f^h_{}(\q{}{},t)\,g\right]+\partial^2_{hk}\left[\varepsilon^{hk}_{}(\q{}{},t)\,g\right]\,,
\end{equation}
includes the drift $\bm{f}(\q{}{},t)$, and the diffusion matrix
\begin{equation}\label{eq:choleski}
\bm{\varepsilon}(\q{}{},t)=\bm{\sigma}^T_{}(\q{}{},t)\,\bm{\sigma}^{}_{}(\q{}{},t)\,.
\end{equation}
This relation can be used to derive the diffusion contribution to the Fokker-Planck operator $\mathscr{L}_{\mathrm{FP}}$ when the It\^o process Eq.~\eqref{eq:sdedyn} is known, as well as to construct the It\^o process when the Fokker-Planck operator is given~\cite{Risken_1996,Frank_2010}.
The second derivation is performed by computing $\bm{\sigma}$ as the Choleski decomposition of the diffusion matrix~$\bm{\varepsilon}$, which is indeed defined by Eq.~\eqref{eq:choleski}.

Alongside the Fokker-Planck forward operator defined in Eq.~\eqref{eq:fpfwop} is possible to introduce the backward operator \cite{Risken_1996}
\begin{equation}\label{eq:fpbwop}
\mathscr{L}_{\mathrm{FP}}^{bw}(\q{}{},t)=f^h_{}(\q{}{},t)\partial^{}_h+\varepsilon^{hk}_{}(\q{}{},t)\partial^2_{hk}\,.
\end{equation}
as the state-space adjoint of the Fokker-Planck forward operator $\mathscr{L}_\mathrm{FP}$, defined by
\mbox{
$
\int g_1\,\mathscr{L}^{}_{\mathrm{FP}}g_2\,\mathrm{d}\q{}{}=\int g_2\,\mathscr{L}^{bw}_{\mathrm{FP}}g_1\,\mathrm{d}\q{}{}\,
$},
where $g_1$ and/or $g_2$ satisfy certain regularity conditions for~\mbox{$|\q{}{}|\rightarrow\infty$}.

The dynamics of the first-order moment $\bm{m}$ is derived from Eq.~\eqref{eq:fpfweq} multiplying by $q^l$, integrating over the \mbox{state-space} $\mathbb{Q}$ and using the definition of the backward operator (\mbox{$g_1=q^l$} and \mbox{$g_2=\rho$}), obtaining
\begin{equation}\label{eq:d1stmomdt}
\dot{m}^{l}_{}(t\,|\,\q{}{0},t_0)=\braket{\mathscr{L}_\mathrm{FP}^{bw}(\q{}{},t)q^l_{}\,|\,\q{}{0},t_0}\,,
\end{equation}
where \mbox{$\braket{g\,|\,\q{}{0},t_0}=\int_\mathbb{Q}g\,\rho(\q{}{},t\,|\,\q{}{0},t_0)\mathrm{d}\q{}{}$} is meant the expectation of $g$ at time $t$ for a distribution that at time $t_0$ ``occupied'' the states $\q{}{0}$, or conditioned to Eq.~\eqref{eq:fpfwic}. Noting that \mbox{$\partial_h q^l=\delta_h^l$} and \mbox{$\partial_{hk}^2q^l=\partial_h \delta_k^l=0$},  equation~(\ref{eq:d1stmomdt}) can be rewritten in terms of the drift
\begin{equation}\label{eq:d1stmomdt2}
\dot{m}^{h}_{}(t\,|\,\q{}{0},t_0)=\braket{f^h_{}(\q{}{},t)\,|\,\q{}{0},t_0}\,,
\end{equation}
and this equation shows that the dynamics of the first-order moment is independent on the diffusion matrix.
An analogous derivation can be performed for computing the dynamics of the covariance, that results
\begin{align}\label{eq:dcovdt}
\dot{s}^{hk}_{}(t\,|\,\q{}{0},t_0)&=\,\braket{f^h_{}(\q{}{},t)(q^k-m^k)\,|\,\q{}{0},t_0}+\nonumber\\
&+\,\braket{f^k_{}(\q{}{},t)(q^h-m^h)\,|\,\q{}{0},t_0}+\nonumber\\
&+\,2\braket{\varepsilon^{hk}_{}(\q{}{},t)\,|\,\q{}{0},t_0}\,.
\end{align}
Expanding in Taylor series up to the first-order the drift and the diffusion matrix
\begin{subequations}\label{eq:taylor1stord}
\begin{align}
f^h(\q{}{},t)&=\,f^h(\bm{m},t)+\partial_l f^h(q^l_{}-m^l_{})\,,\\
\varepsilon^{hk}(\q{}{},t)&=\,\varepsilon^{hk}(\bm{m},t)+\partial_l \varepsilon^{hk}(q^l_{}-m^l_{})\,,
\end{align}
\end{subequations}
and substituting these expansions into Eqs.~\eqref{eq:d1stmomdt2} and \eqref{eq:dcovdt} it has
\begin{subequations}\label{eq:meancovdyn}
\begin{align}
\dot{m}^h_{}(t\,|\,\q{}{0},t_0)&=\,f^h_{}[\bm{m}(t\,|\,\q{}{0},t_0),t]\,,\\
\dot{s}^{hk}_{}(t\,|\,\q{}{0},t_0)&=\,\partial^{}_l f^h_{}[\bm{m}(t\,|\,\q{}{0},t_0),t])\,s^{lk}_{}(t\,|\,\q{}{0},t_0)+\nonumber\\
&+\,\partial^{}_l f^k_{}[\bm{m}(t\,|\,\q{}{0},t_0),t])\,s^{hl}_{}(t\,|\,\q{}{0},t_0)+\nonumber\\
&+\,2\,\varepsilon^{hk}_{}[\bm{m}(t\,|\,\q{}{0},t_0),t]\,,
\end{align}
\end{subequations}
that using vector analysis notation, read as
\begin{subequations}\label{eq:meancovdynvec}
\begin{align}
\bm{\dot m}(t\,|\,\q{}{0},t_0)&=\,\bm{f}[\bm{m}(t\,|\,\q{}{0},t_0^{}),t]\,,\\
\bm{\dot s}&=\,\bm{J}[\bm{m}(t\,|\,\q{}{0},t_0^{}),t]\cdot \bm{s}(t\,|\,\q{}{0},t_0)+\nonumber\\
&+\,\bm{s}(t\,|\,\q{}{0},t_0)\cdot\bm{J}^T_{}[\bm{m}(t\,|\,\q{}{0},t_0^{}),t]+\nonumber\\
&+\,\,2\,\bm{\varepsilon}[\bm{m}(t\,|\,\q{}{0},t_0^{}),t]\,,
\end{align}
\end{subequations}
where $\bm{J}(\bm{m},t)=\partial_{\bm{q}}\bm{f}|_{\bm{m},t}$. The initial condition for this set of ODE is
\begin{subequations}\label{eq:meancovic}
\begin{align}
\bm{m}(t_0^{}\,|\,\q{}{0},t_0^{})&=\,\mathrm{E}[\bm{Q}_0^{}]\,,\\
\bm{s}(t_0^{}\,|\,\q{}{0},t_0^{})&=\,\mathrm{cov}[\bm{Q}_0^{},\bm{Q}_0^{}]\,.
\end{align}
\end{subequations}
Equations \eqref{eq:meancovdyn}, or equivalently Eqs.~\eqref{eq:meancovdynvec}, represent the set of differential equations here briefly addressed as \mbox{\textit{mean-and-covariance dynamics}}, while Eqs.~\eqref{eq:meancovic} are the corresponding initial conditions.

It must be noted that in the case when the drift has an implicit dependence on the parameters, i.e. \mbox{$\bm{f^x}_{}(\bm{x},\bm{p},t)=\bm{f^x}_{}[\bm{x},\bm{g}(\bm{p}),t]$}, the Jacobian of the drift transforms according to
\begin{equation}\label{eq:transfjac}
\partial_{\bm{q}}^{}\bm{f}|_{\bm{m},t}^{}=\bm{\nabla}_{\bm{x},\bm{g}}\bm{f^x}_{}\cdot
\left(\begin{array}{cc}
\bm{I} & \bm{0} \\
\bm{0} & \partial_{\bm{p}}\bm{g}
\end{array}\right)\,,
\end{equation}
this representation is useful when the parameters appear in functions of the device or suspension features, e.g. particle compressibility and density in the acoustophoretic contrast factor.

\subsection{Dynamics of Microparticle Populations}
As stated in the introduction, in order to quantify the separation and dispersion of microparticle populations it is necessary to devise a method that accounts for the statistics of the sample.
The minimum requirement for this model is that it should be able to deal with normally distributed statistics.
This restriction is discussed and amended at the end of this section by using a gaussian mixture to approximate arbitrary PDFs.

As first step, the state-space $\mathbb{Q}$ is split into a coordinate subspace and a parameter subspace, namely \mbox{$\q{}{}=[\bm{x},\bm{p}]^T_{}$} with \mbox{$\bm{x}\in\mathbb{R}_{}^{d_x}=\mathbb{X}$}, \mbox{$\bm{p}\in\mathbb{R}_{}^{d_p}=\mathbb{P}$}, and such that \mbox{$d=d_x+d_p$}.
As a consequence of this splitting, the coordinate marginal and the parameter marginal are
\begin{subequations}\label{eq:marginals}
\begin{align}
\label{eq:xmarginal}\rho^{\bm{x}}_{} (\bm{x},t\,|\,\q{}{0},t_0)&=\,\int_\mathbb{P}\rho(\q{}{},t\,|\,\q{}{0},t_0)\,\mathrm{d}\bm{p}\,,\\
\label{eq:pmarginal}\rho^{\bm p}_{}(\bm{p},t\,|\,\q{}{0},t_0)&=\,\int_\mathbb{X}\rho(\q{}{},t\,|\,\q{}{0},t_0)\,\mathrm{d}\bm{x}\,,
\end{align}
\end{subequations}
respectively, where the coordinate marginal $\rho^{\bm x}_{}$ is the actual distribution of the particle as it is seen in space, while the the parameter marginal $\rho^{\bm p}_{}$ is the distribution over the parameters that is constant in time, i.e. stationary.
Therefore, the drift $\bm{f}(\q{}{},t)$ can be separated into spatial and parameter components
\begin{equation}
\bm{f}(\q{}{},t)=\bm{f}(\bm{x},\bm{p},t)=[\bm{f^x_{}}(\bm{x},\bm{p},t),\bm{f^p_{}}(\bm{x},\bm{p},t)]_{}^T\,.
\end{equation}
Similarly the diffusion matrix can be split as
\begin{equation}
\bm{\varepsilon}(\q{}{},t)=\bm{\varepsilon}(\bm{x},\bm{p},t)=\left(\begin{array}{cc}
\bm{\varepsilon^{xx}_{}}(\bm{x},\bm{p},t) & \bm{0}\\
\bm{0} & \bm{\varepsilon^{pp}_{}}(\bm{x},\bm{p},t)
\end{array}\right)
\end{equation}
where no cross-correlation for the diffusion of particles is allowed between the coordinate subspace and the parameter subspace, i.e. \mbox{$\bm{\varepsilon^{xp}}=\bm{\varepsilon^{px}}=\bm{0}$}. If the particles do not undergo the action of the Brownian motion, that is the case of microparticles, also $\bm{\varepsilon^{xx}}=\bm{0}$.
A convenient choice that constrains the parameter marginal to be constant in time is \mbox{$\bm{f^p_{}}=\bm{0}$} and \mbox{$\varepsilon^{\bm{pp}}_{}=\bm{0}$}. This choice is not unique, for example \mbox{$\bm{f^p_{}}=\bm{m^p_{}}-\bm{p}$} and \mbox{$\bm{\varepsilon^{pp}_{}}=(\bm{\sigma^{pp}_{}})^T\bm{\sigma^{pp}_{}}$} gives the same results in terms of mean-and-covariance dynamics, but reformulates the particle ensemble dynamics in term of SDE instead of ODE. Here, we opt for the ODE form of the particle ensemble dynamics.

With the assumptions so far introduced and dropping SDE notation in favor of ODE notation, Eq.~\eqref{eq:sdedyn} becomes
\begin{subequations}\label{eq:sdemod}
\begin{align}
\label{eq:sdemodx} \bm{\dot X}(t)&=\,\bm{f^x_{}}(\bm{X},\bm{P},t)\,,\\
\label{eq:sdemodp} \bm{\dot P}(t)&=\,\bm{0}\,,
\end{align}
\end{subequations}
and the initial condition is
\begin{subequations}\label{eq:sdemodic}
\begin{align}
\label{eq:sdemodxic} \bm{X}(t_0)&=\,\bm{X}_0\,,\\
\label{eq:sdemodpic} \bm{P}(t_0)&=\,\bm{P}_0\,,
\end{align}
\end{subequations}
with the initial sample such that
\begin{subequations}\label{eq:sdeparic}
\begin{align}
\label{eq:sdeparicx}\bm{X}_0^{}&\sim\,\mathcal{N}(\:\cdot \:|\,\bm{m^x}_0,\bm{s^{xx}}_0)\,,\\
\label{eq:sdeparicp}\bm{P}_0^{}&\sim\,\mathcal{N}(\:\cdot \:|\,\bm{m^p}_0,\bm{s^{pp}}_0)\,,
\end{align}
\end{subequations}
where \mbox{$\mathcal{N}(\:\cdot \:|\bm{m},\bm{s})$} is a multivariate normal distribution with mean $\bm{m}$ and covariance $\bm{s}$. Since, the parameter marginal is time-independent, it is immaterial to write \mbox{$\bm{m^p_{}}$} in place of \mbox{$\bm{m^p}_0$} and the same holds for the variance \mbox{$\bm{s^{pp}}_{}$}.
\begin{figure*}[!!ht]
%\fbox{
\begin{picture}(500,245)
\put(-20,75){\includegraphics[width=6cm]{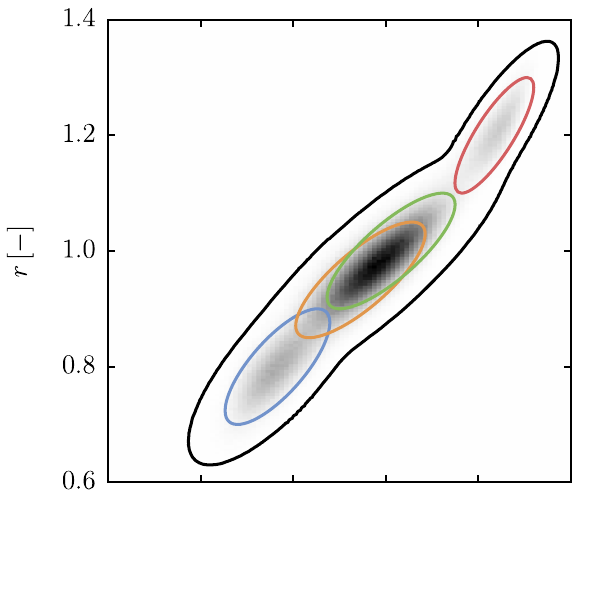}}
\put(-20,0){\includegraphics[width=6cm]{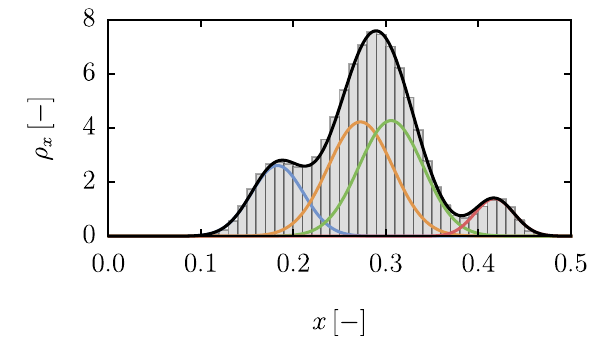}}
\put(150,75.5){\includegraphics[width=3.5cm]{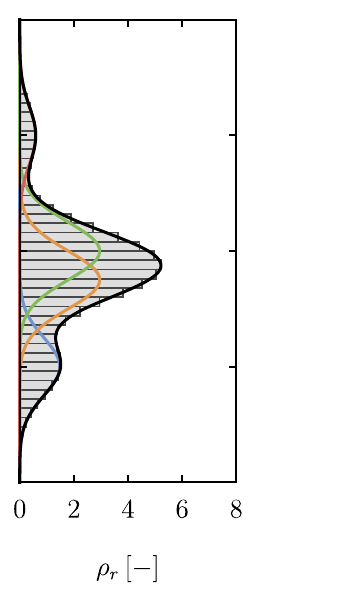}}
\put(245,-3){\includegraphics[width=9cm]{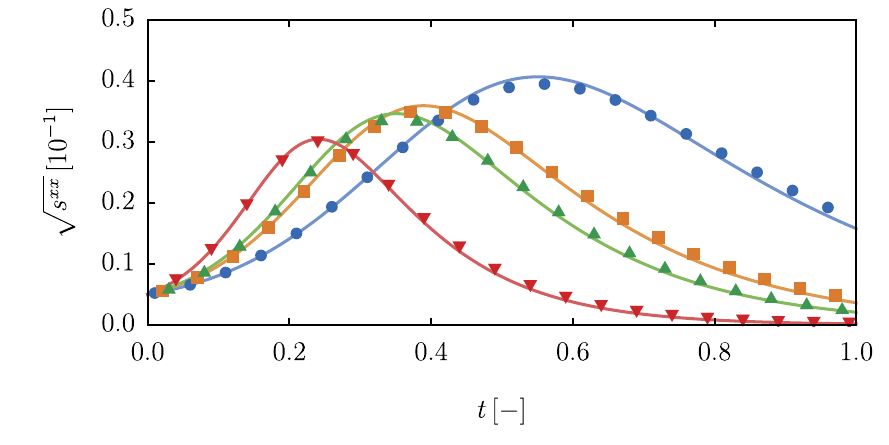}}
\put(245,118){\includegraphics[width=9cm]{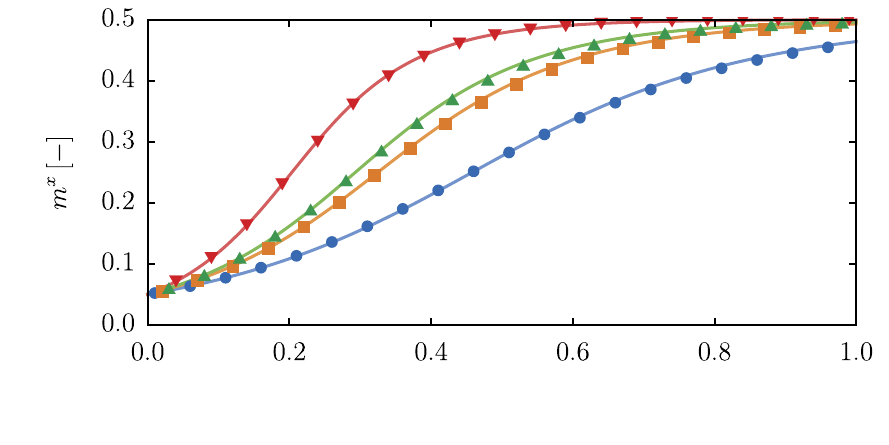}}
\put(12,230){(a)}
\put(12,84){(ax)}
\put(202,230){(ar)}
\put(478,158){(b)}
\put(479,109){(c)}
\end{picture}
%}
\caption{(Color Online) Simulations results for the prototypical model Eq.~\eqref{eq:sdevalid}. Probability density function $\rho(x,r,t)$ (a) from ensemble simulations (grayscale) and approximated by mean-and-covariance dynamics (black line \mbox{$\rho=0.1$}) for four gaussians (colored ellipsis). Spatial marginal \mbox{$\rho^x_{}(x,t)$} (ax) and radius marginal \mbox{$\rho^r_{}(r,t)$} (ar) from ensemble simulations (gray bins) and approximated by mean-and-covariance dynamics (black lines) for four gaussians. Dynamics of the first-order moments $m^x_k$ (b) and the spatial dispersions \mbox{$\sqrt{s^{xx}_{k}}$} (c) for the four kernels (same colors as in panel (a)).}
\label{fig:toymodel}
\end{figure*}
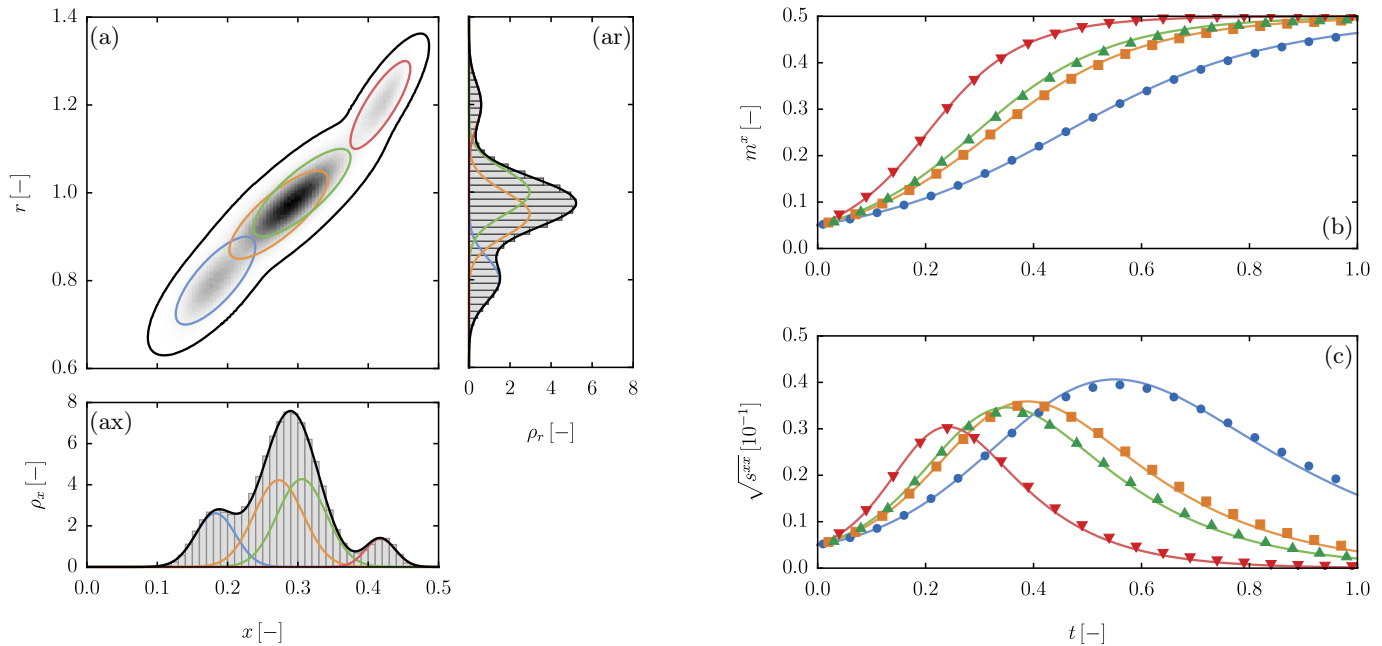
Note that the deterministic process Eq.~\eqref{eq:sdemod} retains the statistics information about the parameter distribution by including a sample $\bm{P}$ that is distributed according to Eqs.~\eqref{eq:sdeparic}.
The mean-and-covariance dynamics associated with Eq.~\eqref{eq:sdemod} can be computed by applying Eq.~\eqref{eq:meancovdynvec} and considering that \mbox{$\bm{\dot m^{p}_{}}=\bm{0}$} and \mbox{$\bm{\dot s^{pp}_{}}=\bm{0}$} because of the stationariety of the parameter marginal (omitting conditionals)
\begin{subequations}\label{eq:meancovdynvecspec}
\begin{align}
\bm{\dot m^x}_{}&=\,\bm{f^x_{}}(\bm{m^x_{}},\bm{m^p_0},t)\,,\\
\bm{\dot s^{xp}_{}}&=\,\partial_{\bm{x}}\bm{f^x_{}}\cdot\bm{s^{xp}_{}}+\partial_{\bm{p}}\bm{f^x_{}}\cdot\bm{s^{pp}}_0\,,\\
\bm{\dot s^{xx}_{}}&=\,2\,\partial_{\bm{x}}\bm{f^x_{}}\cdot\bm{s^{xx}_{}}+2\,\partial_{\bm{p}}\bm{f^x_{}}\cdot\bm{s^{xp}}_{}\,,
\end{align}
\end{subequations}
and the initial condition is given by
\begin{subequations}\label{eq:momcovic}
\begin{align}
\bm{m^x_{}}(t_0\,|\,\q{}{0},t_0^{})&=\,\bm{m^x}_0\,,\\
\bm{s^{xp}_{}}(t_0\,|\,\q{}{0},t_0^{})&=\,\bm{0}\,,\\
\bm{s^{xx}_{}}(t_0\,|\,\q{}{0},t_0^{})&=\,\bm{s^{xx}}_0\,.
\end{align}
\end{subequations}
%Note that Eqs.~\eqref{eq:meancovdynvecspec} are layered, meaning that is possible to solve for the first equation, substituting the result in the second and solve it, finally substitute the result in the third and solve it.

The assumption of normally distributed states can be limiting for describing the PDF, therefore it is proposed to approximate the distribution of the generic population in the state space as a superposition of gaussians,
%\begin{widetext}
\begin{equation}\label{eq:gmrep}
\rho_{}^{}(\q{}{},t\,|\,\q{}{0},t_0^{})=\sum_{k\in\mathcal{K}_{}^{}}w_{k}^{}\,\mathcal{N}\left[\q{}{}\,|\,\bm{m}^{}_{k}(t\,|\,\q{}{0},t_0^{}),\bm{s}^{}_{k}(t\,|\,\q{}{0},t_0^{})\right]\,,
\end{equation}
%\end{widetext}
namely a gaussian mixture model, where $\mathcal{K}$ is a set of gaussians that span the state space, and $w_k^{}$ are weights such that \mbox{$\sum_k w_k=1$}.
Note that adopting this representation allows for the introduction of correlations even in the case when \mbox{$\bm{s^{xp}_{0}}=\bm{0}$}, inasmuch when the gaussians span the state space different weights can be assigned to different locations and thus the PDF exhibits a combined dependence on both $\bm{x}$ and $\bm{p}$.
Extending the representation Eq.~\eqref{eq:gmrep} in the case of multiple populations and considering the initial parameter configuration independent on the initial spatial positions, the parameter marginal for the $h$-th population can be approximated as
\begin{equation}\label{eq:kmpar}
\rho^{\bm{p}}_h(\bm{p})=\sum_{k\in\mathcal{K}_h}w_{hk}^{}\,\mathcal{N}(\bm{p}\,|\,\bm{m^p}_{hk},\bm{s^{pp}}_{hk})\,,
\end{equation}
where $w_{hk}$, $\bm{m^p}_{hk}$ and $\bm{s^{pp}}_{hk}$ are the weight, the means and the covariance associated with the $k$-th gaussian in the $h$-th population (in the following the subscript ``$h$'' is meant to address the population, while the subscript ``$k$'' refers to the gaussian).
The solution of Eqs.~\eqref{eq:meancovdynvecspec} allows then to approximate the spatial marginal for the $h$-th population
\begin{equation}\label{eq:kmspace}
\rho^{\bm{x}}_h(\bm{x},t\,|\,\q{}{0},t_0^{})=\sum_{k\in\mathcal{K}_h}w_{hk}\,\mathcal{N}[\bm{x}\,|\,\bm{m^x}_{hk}(t\,|\,\q{}{0},t_0^{}),\bm{s^{xx}}_{hk}(t\,|\,\q{}{0},t_0^{})]\,.
\end{equation}
%so that by using this reconstruction and the knowledge of the operative conditions is possible to derive the performance indicators as described Sec.~\ref{sec:examples}~(C).
%The price of evolving a moderate number of mean-and-covariance equations, e.g. $\mathrm{dim}(\mathcal{K})\sim 100$, is still comparably smaller than that necessary to solve the direct numerical simulations for which $N\sim 10^5-10^6$.

Equations~\eqref{eq:meancovdynvecspec}, the approximation~\eqref{eq:kmpar}, and the reconstruction~\eqref{eq:kmspace} forms the MCDGM method that applied to acoustphoresis models is used to approximate the dynamics of microparticle populations during acoustophoretic separation and to derive the separation indicators.

\section{Examples}\label{sec:examples}

\subsection{Minimal Working Model}
In order to illustrate the basic features of the MCDGM method when applied to acoustophoresis, we consider an one-dimensional prototypical model for a single particle population with a radius distribution.
Therefore, we can assume that the ensemble dynamics is given by  (\mbox{$h=1$})
\begin{equation}\label{eq:sdevalid}
\dot{X}(t)=R_{}^2\sin[2\pi\,X(t)]\,,\\
\end{equation}
where $X$ are the particle positions, and $R$ are corresponding the particle radii.
The initial conditions and the radius distribution are generated by the weighted superposition of four normally distributed random processes
\begin{subequations}\label{eq:sdevalidic}
\begin{align}
X_k^{}(0)&\sim\,\mathcal{N}(\:\cdot \:|\,m^x_{0,k},s^{xx}_{0,k})\,,\\
R_k^{}&\sim\,\mathcal{N}(\:\cdot \:|\,m^r_{k},s^{rr}_{k})\,,
\end{align}
\end{subequations}
with spatial averages $m_{0,k}^x$ and variances $s_{0,k}^{xx}$, and with radius averages $m_{k}^r$ and variances $s_{k}^{rr}$.
Note that once the initial distribution is generated, Eq.~\eqref{eq:sdevalid} does not retain any information about the four processes that have generated the ensemble.
However in the particle ensemble simulations, a \textit{tag} corresponding to the generating process is applied to the particle to reconstruct the dynamics of the statistics for the single gaussians.

The mean-and-covariance dynamics for the $k$-th gaussian associated with Eq.~\eqref{eq:sdevalid} read as
\begin{subequations}\label{eq:meancovmodel}
\begin{align}
\dot{m}^x_k&=\,(m_{k}^r)^2\,\sin(2\pi\,m^x_k)\,,\\
\dot{s}^{xr}_k&=\,2\pi\,(m_{k}^r)^2\,\cos(2\pi\,m^x_k)\,s^{xr}_k+\nonumber\\
&+\,2\,m_{k}^r\,\sin(2\pi\,m^x_k)\,s^{rr}_{k}\,,\\
\dot{s}^{xx}_k&=\,4\pi\,(m_{k}^r)^2\,\cos(2\pi\,m^x_k)\,s^{xx}_k+\nonumber\\
&+4\,m_{k}^r\,\sin(2\pi\,m^x_k)\,s^{xr}_k\,.
\end{align}
\end{subequations}
The numerical solutions of this equations are thus compared with ensemble simulations with $10^5$ total particles generated by Eq.~\eqref{eq:sdevalidic} with un-normalized weights \mbox{$w=\{0.5, 1.0, 1.0, 0.2\}$}, all starting at position \mbox{$m^x_{}(0)=5\cdot 10^{-2}$} with spatial standard deviation \mbox{$\sigma^{xx}_{}(0)=5\cdot 10^{-3}$}. Four average radius were considered \mbox{$m^r_{}=\{0.8,0.95,1.0,1.2\}$} with standard deviation \mbox{$\sigma^{rr}=5\cdot 10^{-2}$}, and cross-covariance \mbox{$\sigma^{xr}(0)=0$}. Integration of Eq.~\eqref{eq:sdevalid} and Eq.~\eqref{eq:meancovmodel} were performed by the Matlab routine \verb|ode45| with suitable options as to ensure convergence and accuracy.

Figure~\ref{fig:toymodel} reports the comparisons of the direct numerical simulations for the particle ensemble and the mean-and-covariance dynamics for the four gaussians.
Panel (a) compares the probability density function in the state space at \mbox{$t=0.35$} for the ensemble simulations (grayscale) and for the gaussian mixture:
the black line is the isolevel at \mbox{$\rho^{}_{}(x,r,t)=0.1$}, while the colored ellipsis are the confidence ellipsis for the four gaussians.
Panel (ax) plots the coordinate marginal \mbox{$\rho^x_{}(x,t)$} for the particle-ensemble simulations (gray bins) and that reconstructed by the gaussian mixture (black line). The colors for the gaussians correspond to the colors for the confidence ellipsis in panel (a).
Finally, panel (ar) shows the radius-marginal \mbox{$\rho^r_{}(r,t)$} that is stationary in time and corresponds to the radius distribution.
As it can see, the MCDGM method provides a good approximation for the four gaussians as well as the PDF.
This can be appreciated by the quantitative comparisons in \mbox{Fig.~\ref{fig:toymodel}(a)-(b)} where
the average population positions, namely the first-order moments \mbox{$m^x_{}(t)$} (panel a), and the spatial dispersion \mbox{$\sqrt{s^{xx}_{}}$} (panel b) for the
four gaussians are plotted as function of the time. The symbols correspond to the data extracted from the particle-ensemble simulations by using the particle tags,
while the lines are computed by the MCDGM method.
In all of the four cases, MCDGM provides a good approximation of the statistics for the marginals of the four gaussians, and as a consequence of the spatial marginal.

\subsection{Buffer-Dependent Separation Performance in Acoustophoresis of Blood Components}\label{sec:SIPmodel}
\begin{figure}[!!t]
%\fbox{\begin{picture}(226,148)
%\fbox{
\begin{picture}(226,149)
\put(-40,-15){\includegraphics[width=10cm]{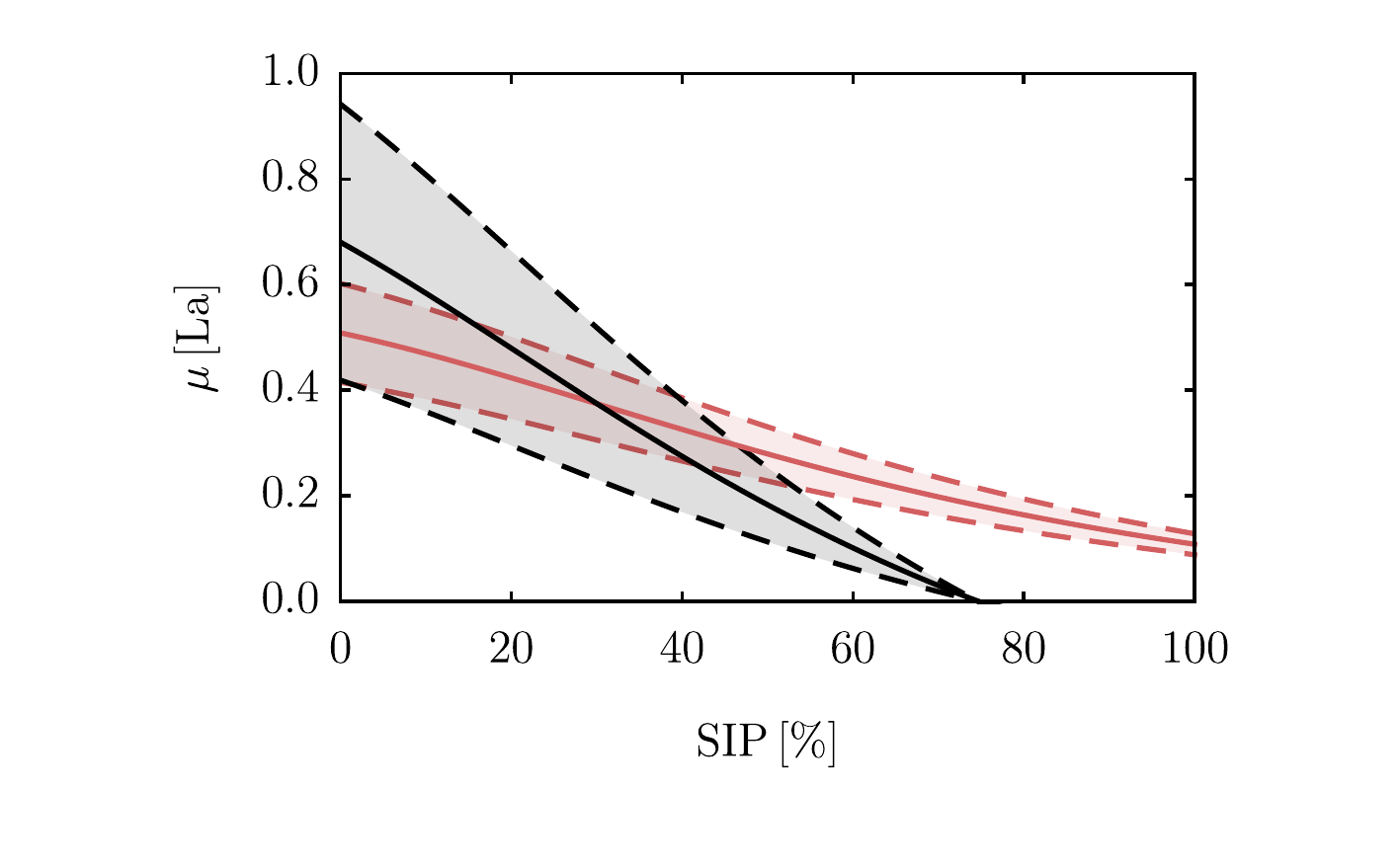}}
\end{picture}
%}
\caption{(Color Online) Mobility of the RBCs (red) and the WBCs (black) as function of the SIP concentration \mbox{$\mathrm{SIP}\%$}. Continuous line is the mobility computed for the average radius $m^r_h$ of the particle populations by Eq.~\eqref{eq:mobility}, while the areas indicate the ranges \mbox{$m^r_h\pm \sigma^r_h$}. (\mbox{$\mathrm{La}[=]10^{-9}\si{m^3_{}\,s^{-2}_{}\,kg^{-1}_{}}$})
}
\label{fig:rbcwbcmob}
\end{figure}

Adjustments in the carrier-fluid properties to enhance the separation performance has been successfully employed in acoustophoretic separation involving diluted blood samples~\cite{urbansky2017rapid}.
The authors, in place of using pure Phosphate Buffer Saline (PBS) in which RBCs/WBCs separation was highly unefficient, employed PBS and Stock Isotonic Percoll (SIP) at different dilution rates (\mbox{$\mathrm{SIP}\%$}) to change the fluid properties and consequently the acoustic contrast factor for both the RBCs and the WBCs.
Because of the specific cell properties, as the concentration of SIP is increased the mobility of WBCs decreases as much as that of RBCs, see Fig.~\ref{fig:rbcwbcmob}.
About \mbox{$\mathrm{SIP}=30\%$} the mobility of WBC and RBC are almost equivalent, and in the correspondence of \mbox{$\mathrm{SIP}\%\approx 70\%$} the acoustic mobility for the WBC population approaches zero and thus the two populations can be successfully separated being the WBCs segregated in the correspondence of the inlet position.
In the following, these experiments are simulated by considering a 1D model of separation, radius distribution derived from Coulter Counter measurements, and density and compressibility measurements adapted from~\cite{cushing2017ultrasound}.

\begin{figure}[!!t]
%\fbox{\begin{picture}(226,148)
%\fbox{
\begin{picture}(226,283)
\put(-40,-15){\includegraphics[width=10cm]{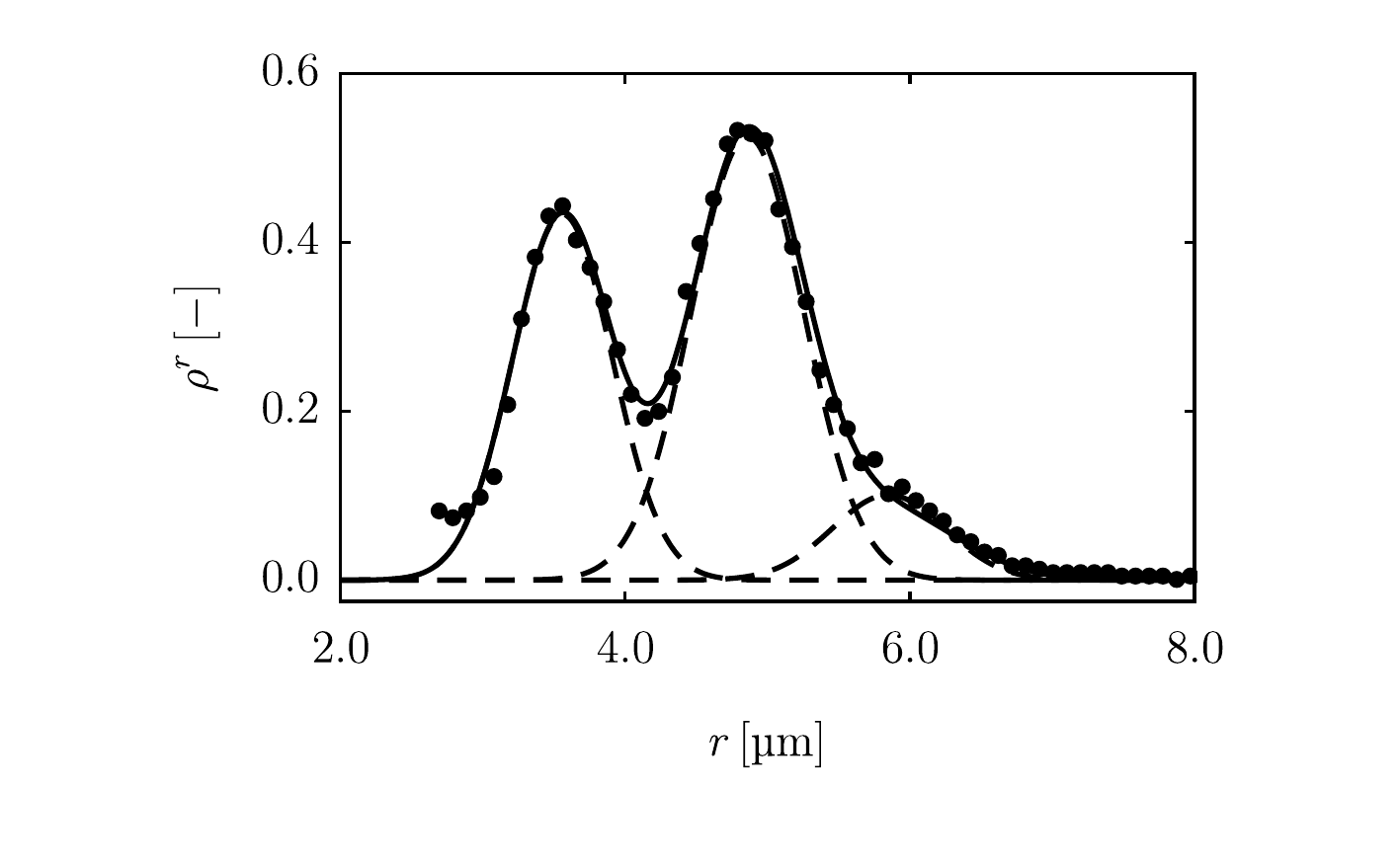}}
\put(-40,120){\includegraphics[width=10cm]{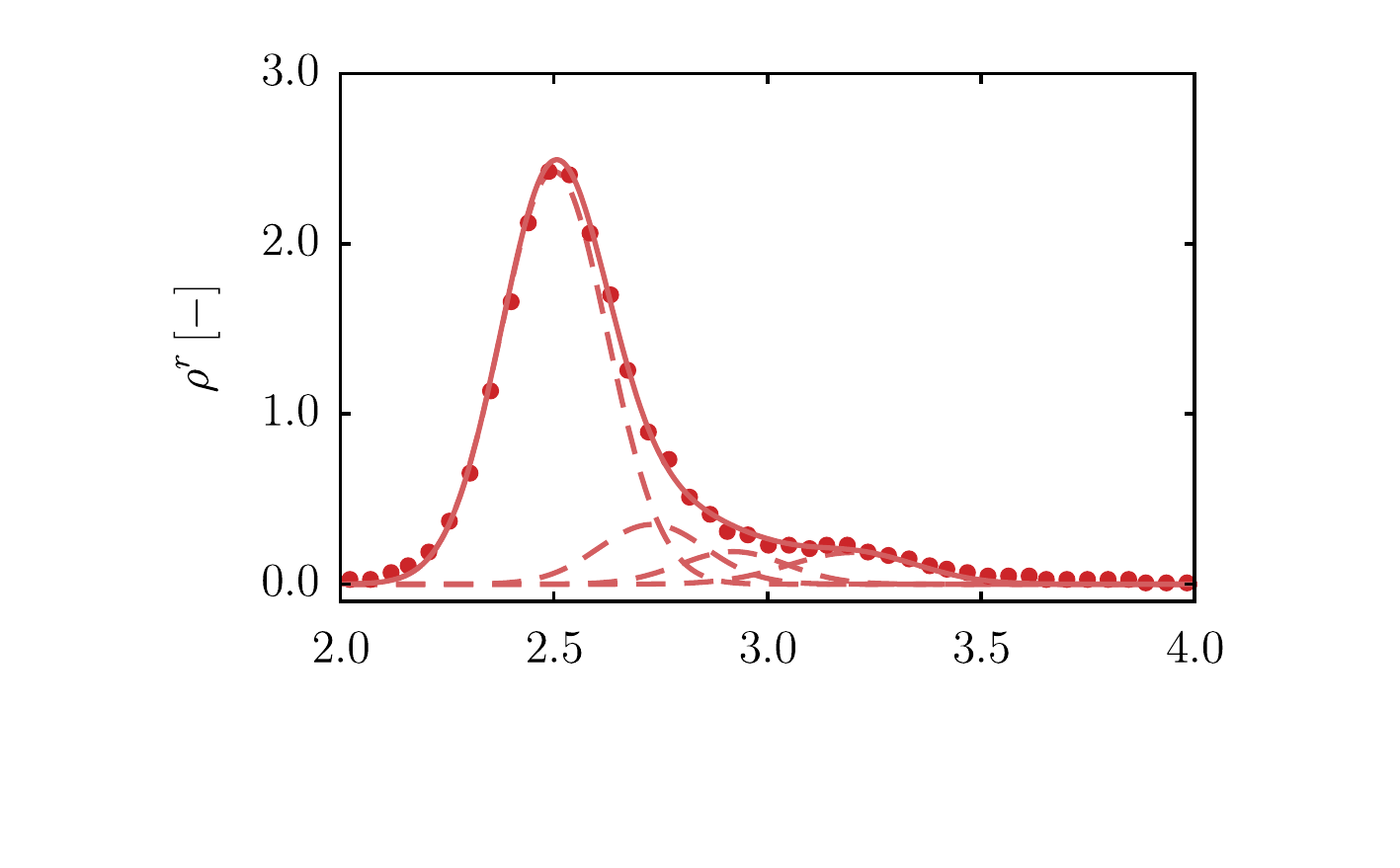}}
\put(173,270){(RBC)}
\put(170,135){(WBC)}
\end{picture}
%}
\caption{(Color Online) Radius distribution from Coulter Counter (symbols) and approximated by gaussian-mixtures (black) for RBC and WBC populations.
The gaussian-mixture parameters and particle physical parameters are:
}
\label{fig:histograms}
\begin{ruledtabular}
\begin{tabular}{lcccccc}
  h & $k$ & $\kappa\,[\si{TPa^{-1_{}}}]$ & $\rho\,[\si{kg\,m ^{-3}_{}}]$ & $w_{hk}^{}$ &$m^r_{hk}\,[\si{\micro m}]$ & $\sigma^{rr}_{hk}\,[\si{\micro m}]$ \Tstrut \\
\hline
 RBC & $1$ & $334$ & $1101$ & $0.76$ &  $2.50$ & $0.125$ \Tstrut\\
 & $2$ & $334$ & $1101$ & $0.11$ &  $2.73$ & $0.125$ \Tstrut\\
 & $3$ & $334$ & $1101$ & $0.06$ &  $2.92$ & $0.125$ \Tstrut\\
 & $4$ & $334$ & $1101$ & $0.07$ &  $3.20$ & $0.125$ \Tstrut \\
\hline
 WBC & $1$ & $393$ & $1054$ & $0.38$ &  $3.56$ & $0.350$ \Tstrut\\
 & $2$ & $393$ & $1054$ & $0.52$ &  $4.86$ & $0.395$ \Tstrut\\
 & $3$ & $393$ & $1054$ & $0.10$ &  $5.85$ & $0.500$ \Tstrut
\end{tabular}
\end{ruledtabular}
\end{figure}

The 1D model equation with radius distribution read as 
\begin{equation}\label{eq:ankemodel}
\dot{Y}_h^{}(t)=\epsilon\,\mu(R_h^{},\tilde\kappa_h^{},\tilde\rho_h^{},\eta_\mathrm{f}^{})\,\sin\left[\frac{2\pi\,Y_h^{}(t)}{W}\right]\,,
\end{equation}
with \mbox{$h=\mathrm{RBC}\,,\mathrm{WBC}$}. In this equation, \mbox{$\epsilon=\pi\,E_\mathrm{ac}^{}/W$} where \mbox{$E_\mathrm{ac}^{}=10\,\si{J\,m^{-3}_{}}$} is the acoustic energy density, and \mbox{$W=375\,\si{\micro m}$} is the channel width. The acoustophoretic mobility is given by
\begin{equation}\label{eq:mobility}
\mu(r,\tilde\kappa,\tilde\rho,\eta_\mathrm{f}^{})=
\frac{2\,\Phi(\tilde\kappa,\tilde\rho)\,r^2_{}}{9\,\eta_\mathrm{f}^{}}\,,
\end{equation}
where $r$ is the particle radius, $\eta_\mathrm{f}^{}$ is the fluid viscosity and the contrast factor
\begin{equation}\label{eq:acf}
\Phi(\tilde\kappa,\tilde\rho)=(1-\tilde\kappa)+\frac{3}{2}\frac{2\,(\tilde\rho-1)}{2\,\tilde\rho+1}\,,
\end{equation}
is a function of the particle/fluid compressibility ratio \mbox{$\tilde\kappa=\kappa/\kappa_\mathrm{f}^{}$},
and the  particle/fluid density ratio \mbox{$\tilde\rho=\rho/\rho_\mathrm{f}^{}$}.
The fluid compressibility and density are considered as function of the SIP concentration using the polynomial interpolations described in~\cite{urbansky2017rapid}.
The radius distributions for the RBCs and the WBCs are given in terms of gaussian mixtures
\begin{equation}
R_{hk}^{}\sim\,\mathcal{N}(\:\cdot \:|\,m^r_{hk},s^{rr}_{hk})\,,\quad h=\mathrm{RBC}\,,\mathrm{WBC}\,,
\end{equation}
which are shown in Fig.~\ref{fig:histograms} and for which the caption reports the gaussian mixture parameters and the physical parameters.
The initial spatial distributions for the two cell types are
\begin{equation}
Y_{hk}^{}(0)\sim\,\mathcal{N}(\:\cdot \:|\,m^y_{0,hk},s^{yy}_{0,hk})\,,\quad h=\mathrm{RBC}\,,\mathrm{WBC}\,,
\end{equation}
where $m^y_{0,hk}=0.05\,W$ and $s^{xx}_{0,hk}=(5\times 10^{-3}\,W)^2_{}$, meaning that they have the same starting position and the initial spread.

\begin{figure*}[!!ht]
%\fbox{
\begin{picture}(500,570)
\put(-40,-20){\includegraphics[width=10cm]{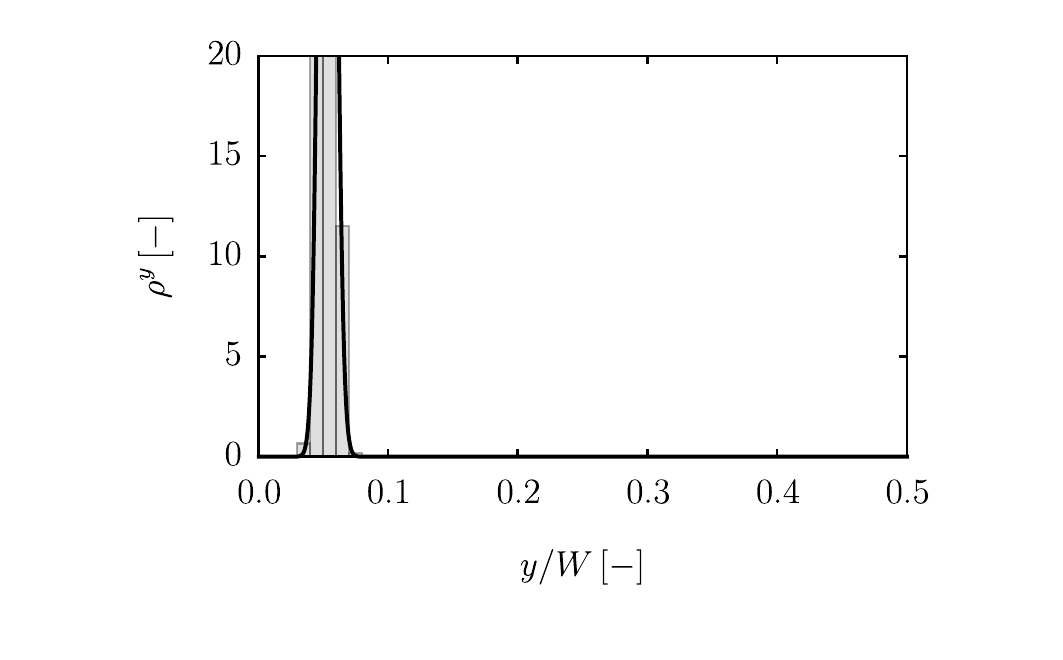}}
\put(-40,120){\includegraphics[width=10cm]{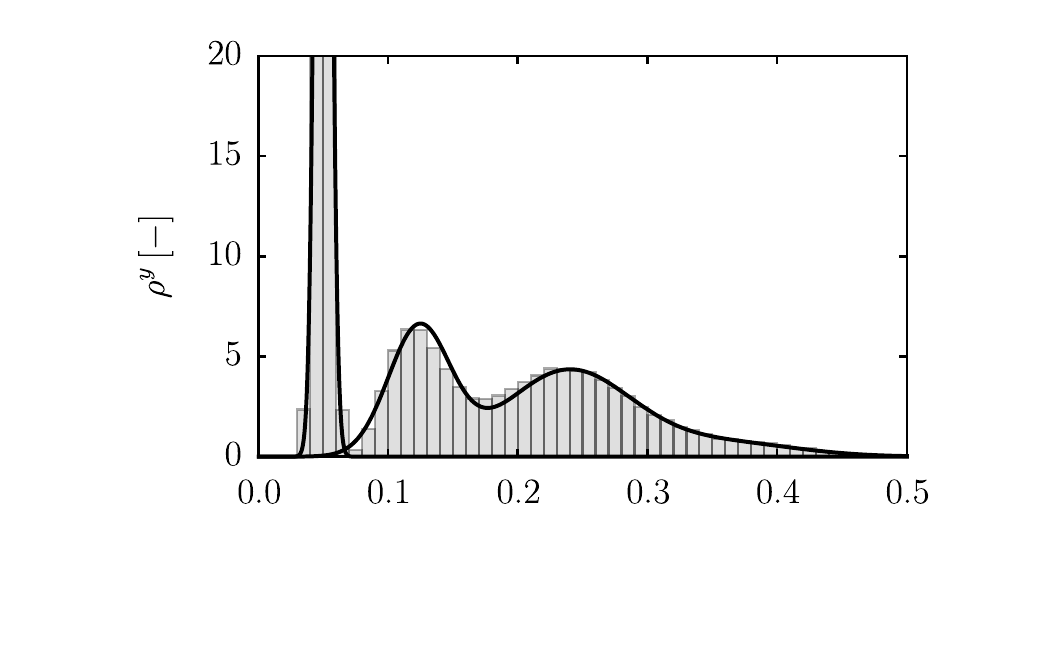}}
\put(-40,260){\includegraphics[width=10cm]{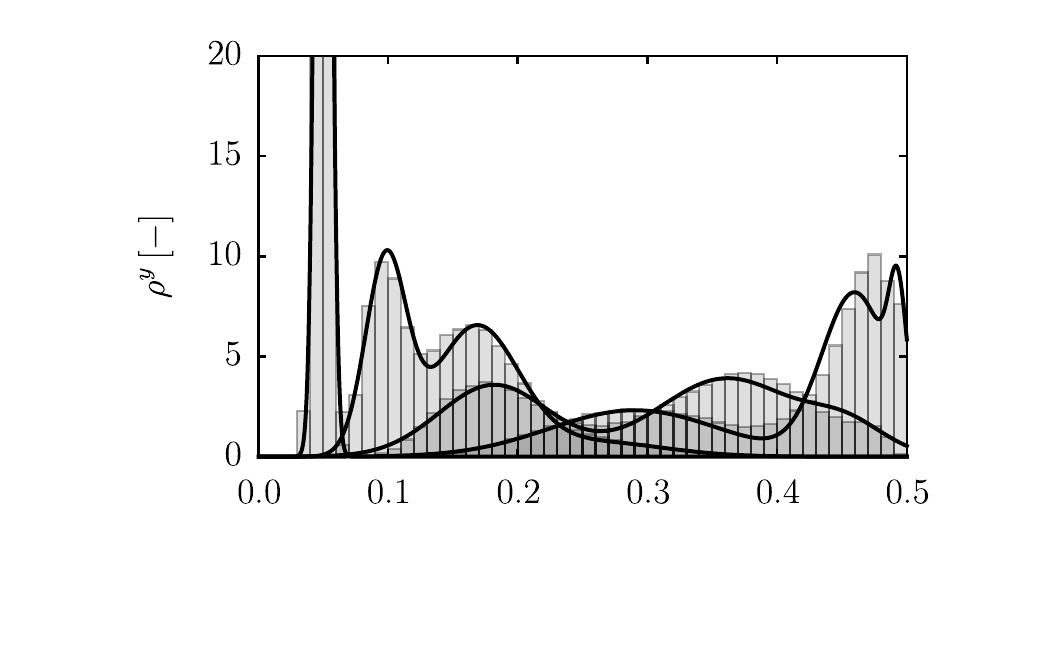}}
\put(-40,400){\includegraphics[width=10cm]{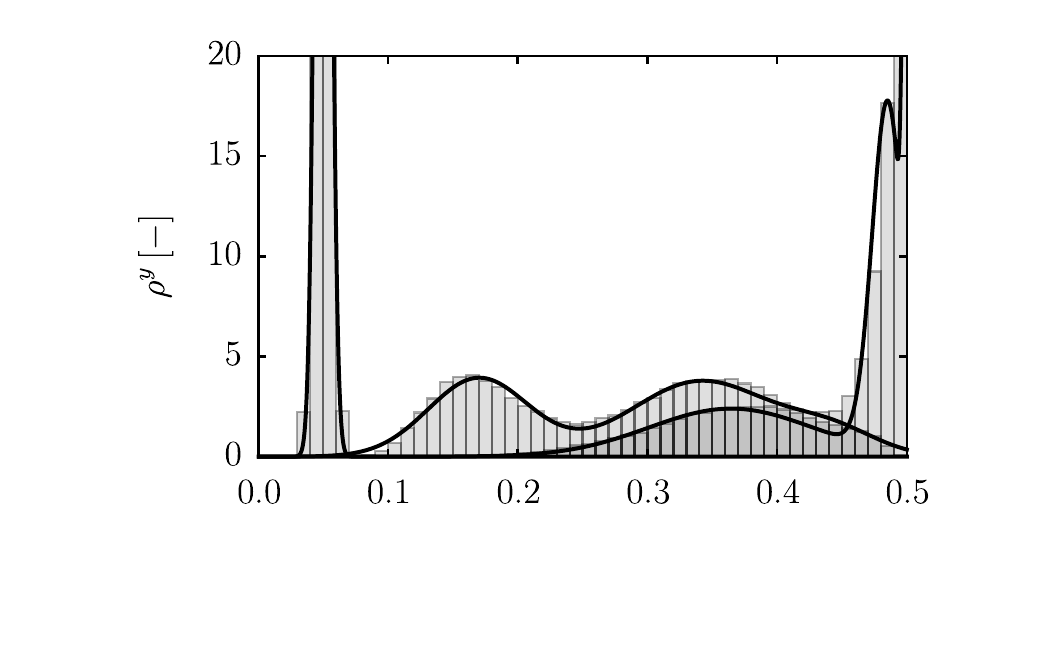}}
\put(230,-20){\includegraphics[width=10cm]{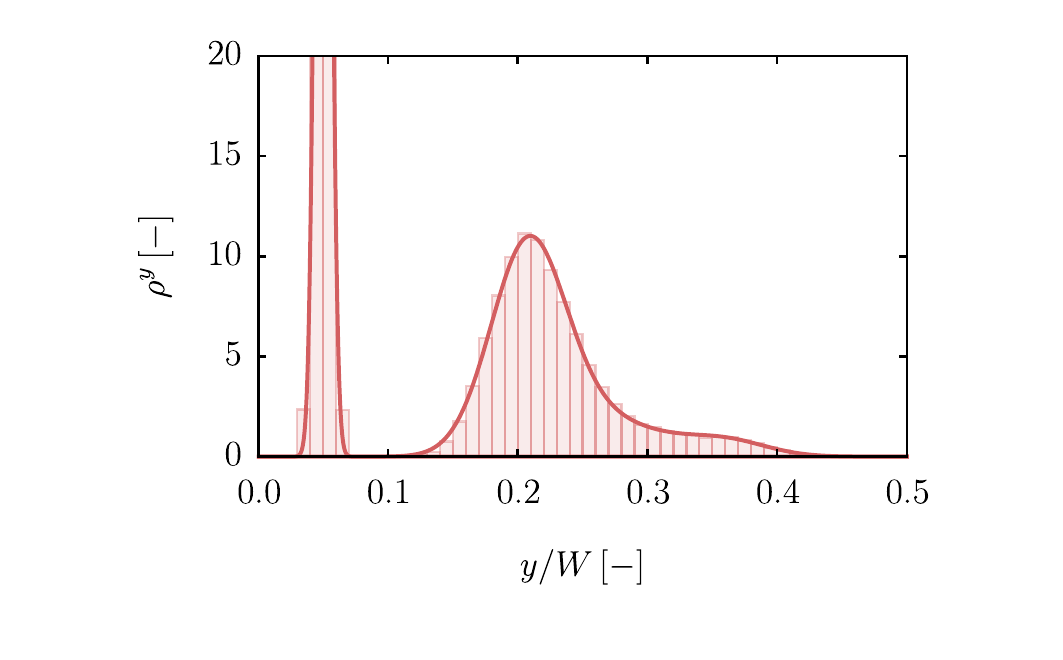}}
\put(230,120){\includegraphics[width=10cm]{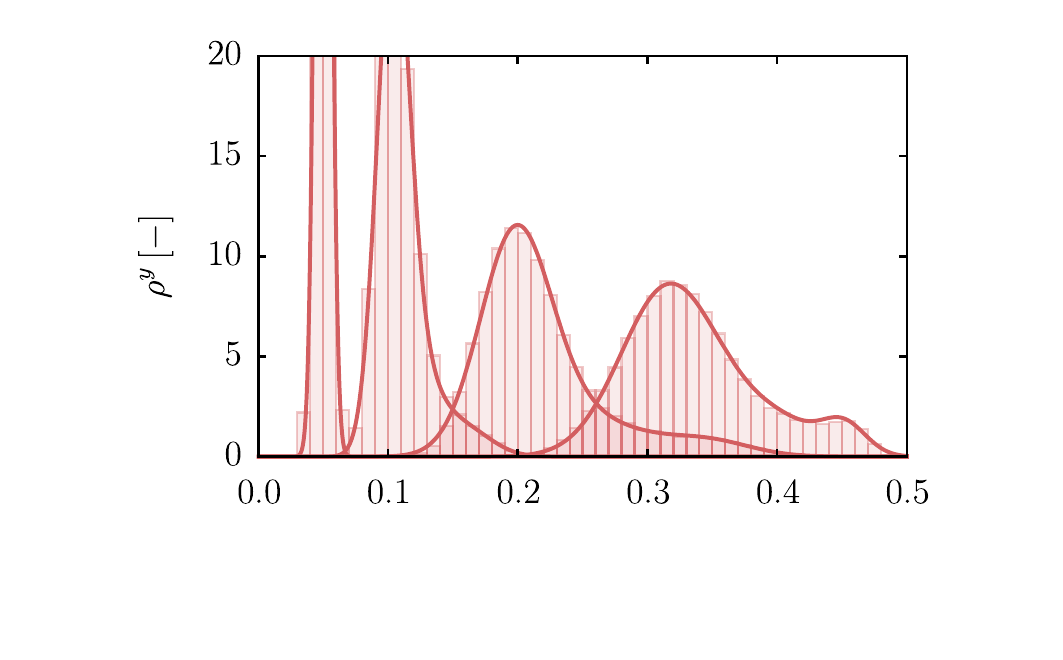}}
\put(230,260){\includegraphics[width=10cm]{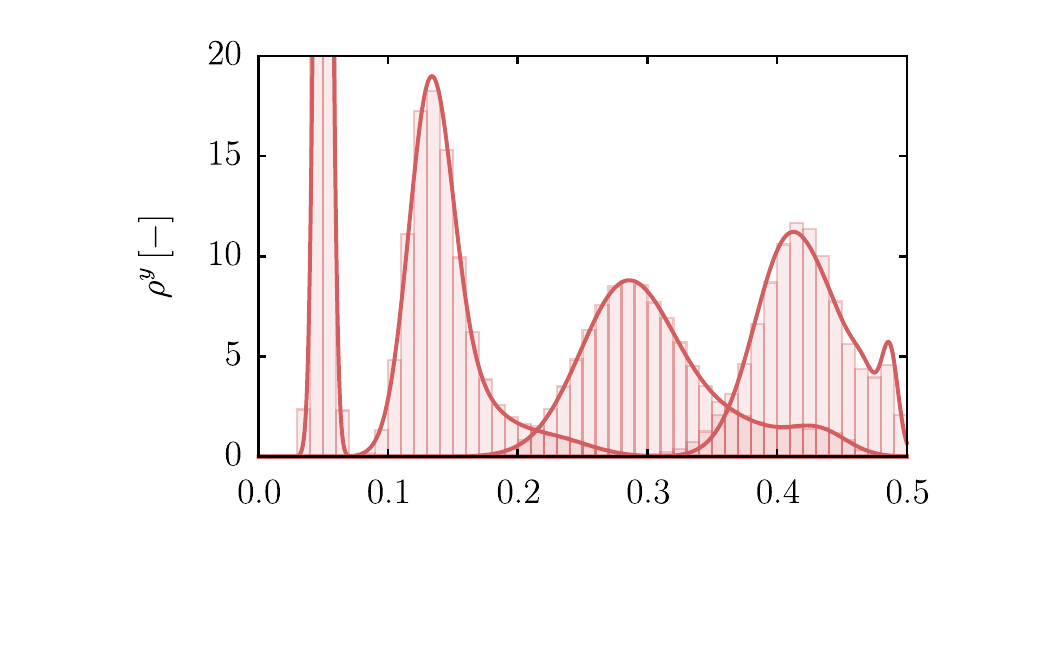}}
\put(230,400){\includegraphics[width=10cm]{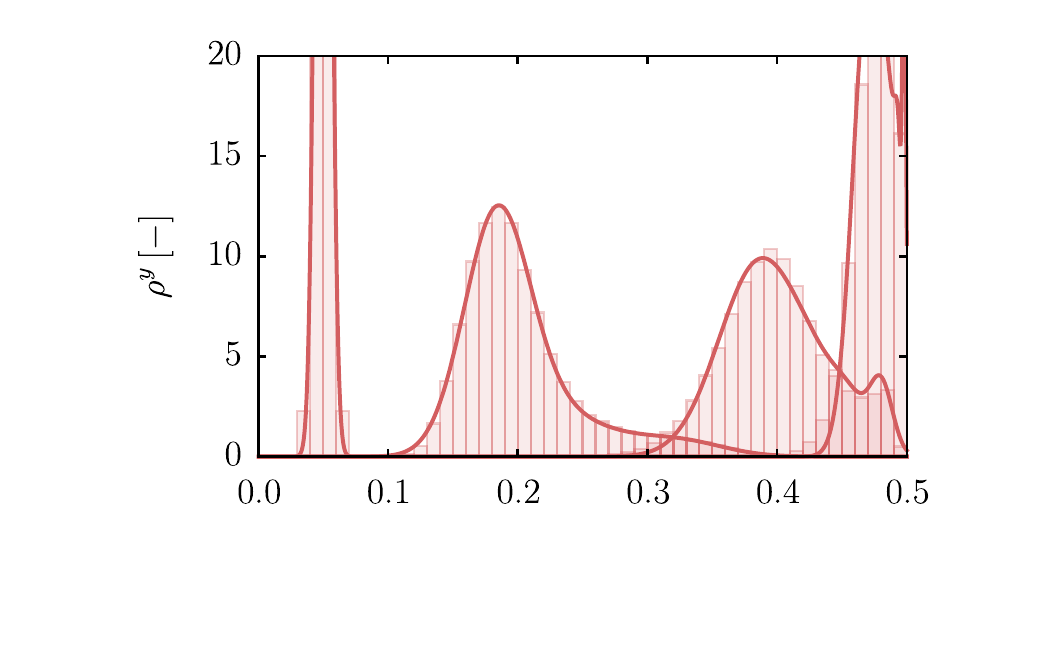}}
\put(53,540){$t=0$}\put(80,478){$t=2$}\put(173,540){$t=4$}
\put(53,400){$t=0$}\put(78,353){$t=2$}\put(150,340){$t=4$}\put(180,371){$t=6$}
\put(53,260){$t=0$}\put(102,200){$t=6$}
\put(53,120){$t\leq 6$}
\put(323,540){$t=0$}\put(354,525){$t=2$}\put(427,512){$t=4$}\put(436,540){$t=6$}
\put(353,395){$t=2$}\put(389,365){$t=4$}\put(434,379){$t=6$}
\put(344,255){$t=2$}\put(359,239){$t=4$}\put(400,224){$t=6$}
\put(323,120){$t=0$}\put(362,96){$t=6$}
\put(205,560){(L1)}
\put(475,560){(R1)}
\put(205,420){(L2)}
\put(475,420){(R2)}
\put(205,280){(L3)}
\put(475,280){(R3)}
\put(205,140){(L4)}
\put(475,140){(R4)}
\end{picture}
%}
\caption{(Color Online) Simulations results for the WBC/RBC separation model Eq.~\eqref{eq:ankemodel} with population distribution given in Fig.~\ref{fig:histograms}.
WBC (left) and RBC (right) $y$-marginal from ensemble simulations (bins) and MCDGM (lines) at different times as the $\mathrm{SIP}$ concentration varies:
\mbox{$\mathrm{SIP}=0\%$} (L1-R1), \mbox{$\mathrm{SIP}=30\%$} (L2-R2), \mbox{$\mathrm{SIP}=50\%$} (L3-R3), \mbox{$\mathrm{SIP}=70\%$} (L4-R4).
}
\label{fig:WBCRBCcomparison}
\end{figure*}
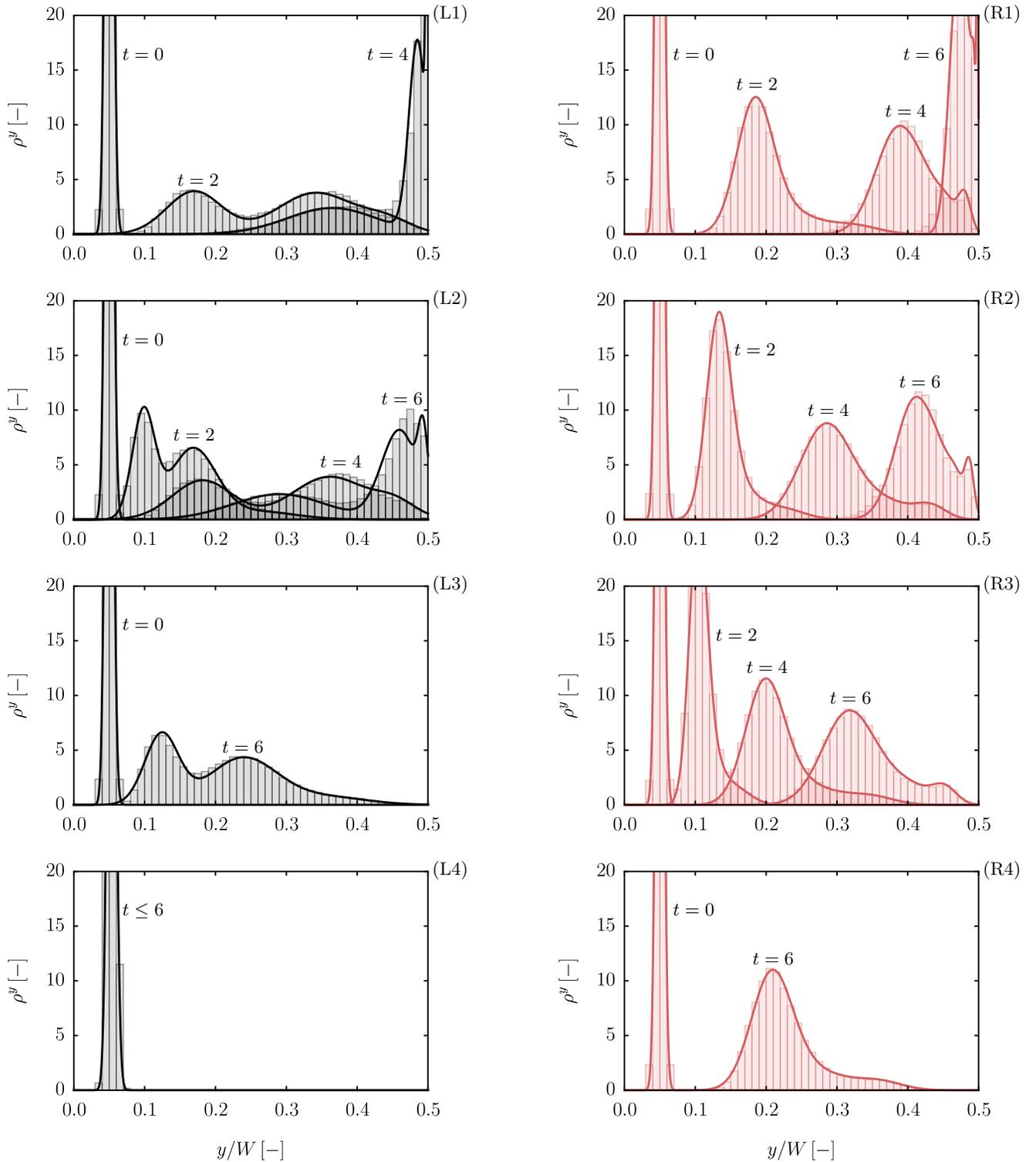

The MCDGM equations corresponding to the ensemble Eq.~\eqref{eq:ankemodel} can be derived by Eqs.~\eqref{eq:meancovdynvecspec} applying the transformation Eq.~\eqref{eq:transfjac}. The equations obtained are formally identical to Eqs.~\eqref{eq:meancovmodel}
\begin{subequations}\label{eq:WBCRBCmodel}
\begin{align}
\dot{m}^y_{hk}&=\,\epsilon\,\mu_{hk}^{}\,\sin\left(\frac{2\pi\,m^y_{hk}}{W}\right)\,,\\
\dot{s}^{yr}_{hk}&=\,\epsilon\,\mu_{hk}^{}\,\frac{2\pi}{W}\,\cos\left(\frac{2\pi\,m^y_{hk}}{W}\right)\,s^{yr}_{hk}+\nonumber\\
&+\epsilon\,\mu'_{hk}\,\sin\left(\frac{2\pi\,m^y_{hk}}{W}\right)\,s^{rr}_{hk}\,,\\
\dot{s}^{yy}_{hk}&=\,2\,\epsilon\,\mu_{hk}^{}\frac{2\pi}{W}\cos\left(\frac{2\pi\,m^y_{hk}}{W}\right)\,s^{yy}_{hk}+\nonumber\\
&+2\,\epsilon\,\mu'_{hk}\,\sin\left(\frac{2\pi\,m^y_{hk}}{W}\right)\,s^{yr}_{hk}\,.
\end{align}
\end{subequations}
where \mbox{$\mu_{hk}^{}=\mu(m^r_{hk},\tilde\kappa_h^{},\tilde\rho_h^{},\eta_\mathrm{f}^{})$} is the mobility and \mbox{$\mu'_{hk}=\partial_r^{}\mu(m^r_{hk},\tilde\kappa_h^{},\tilde\rho_h^{},\eta_\mathrm{f}^{})$} is the derivative of the mobility with respect to the radius calculated, both calculated for the average radius $m_{hk}^r$.
The integration of Eq.~\eqref{eq:ankemodel} and Eqs.~\eqref{eq:WBCRBCmodel} were performed by using the Matlab routine \verb|ode45|  with suitable parameters to ensure convergence and accuracy.

Figure \ref{fig:WBCRBCcomparison} shows the comparison of spatial marginal $\rho^y_{}$ resulting from particle-ensemble simulations (bins) and the MCDGM method (lines) for WBC (black) and RBC (red) for different times and at different SIP concentrations.
For pure PBS buffer, i.e. \mbox{$\mathrm{SIP}=0\,\%$}, the mobility of WBCs is higher than that of RBCs and the two gaussians corresponding to the largest radii of the WBCs already moved at the channel center-line for \mbox{$t=4$}. At the same time-instant the RBCs, to which the gaussian for the smallest radius contributes for the largest part, are still located at $y/W\simeq 0.4$ or better in the range \mbox{$0.3<y/W<0.5$}.
For \mbox{$\mathrm{SIP}=30\,\%$}, there is not an appreciable difference between the mobility of the WBCs and that of the RBCs (see Fig.~\ref{fig:rbcwbcmob}), and for \mbox{$t=6$} the gaussians for the two largest WBC radii are located at \mbox{$0.4<y/W<0.5$}, while the largest part of RBCs occupy the region \mbox{$0.35<y/W<0.5$}.
For \mbox{$\mathrm{SIP}=50\,\%$}, there is a dramatic change in mobility for WBCs that now for \mbox{$t=6$} are located in the range \mbox{$0.05<y/W<0.4$}, while
the RBCs are in the region \mbox{$0.2<y/W<0.5$}.
For \mbox{$\mathrm{SIP}=70\,\%$}, the WBCs reaches the isoacoustic concentration so that they remain close to the initial point for times \mbox{$t\leq 6$}. The RBCs still have an appreciable mobility for this SIP concentration and for \mbox{$t=6$} they occupy the region \mbox{$0.1<y/W<0.4$}.
Finally, also in this case one can appreciate the good approximation properties of the MCDGM method when compared with the ensemble simulations for microparticle distributions and mobility values occurring in real-world applications.

\subsection{Free-Flow Acoustophoretic Separation}
So far, although the approximation properties of the MCDGM method have been illustrated, the method has not been applied to any model corresponding to a real-world case of acoustophoretic separation, namely free-flow acoustophoretic separations.
In order to do this, (i)~one needs to introduce a model for the axial flow that takes into account for the hydrodynamics parameters, such as the overall flowrate and the side/center flowrate ratios, and (ii)~it is necessary to develop further the MCDGM method to introduce the separation indicators. 
In this section we investigate the approximation properties of the MCDGM method and study the reliability of the method when some of the separation parameters vary while adopting different prefocusing strategies.
\paragraph*{Axial Flow Model.} 
Let us consider the inlet flowrate ratio
\begin{equation}
q_\mathrm{in}^{}=\frac{Q_{\mathrm{s,in}}^{}}{Q_{\mathrm{c,in}}^{}}\,,
\end{equation}
where $Q_{\mathrm{s,in}}^{}$ is the inlet flowrate at sides and $Q_{\mathrm{c,in}}^{}$ is the inlet flowrate at the center, \mbox{$Q=Q_{\mathrm{s,in}}^{}+Q_{\mathrm{c,in}}^{}$} is then the total flowrate.
Given the flowrate ratio \mbox{$q_\mathrm{in}^{}$}, it is possible to estimate the position of the streamline separating the side and the center inlet streams \mbox{$y_\mathrm{fj}^{}$} (``fj'' stands for flow-joining) by assuming
\begin{equation}\label{eq:qinyfj}
q_\mathrm{in}^{}(y_\mathrm{fj}^{})=\frac{2\int_0^{y_\mathrm{fj}^{}}\int_0^H u_\mathrm{ax}(y,z)\,\mathrm{d}y\mathrm{d}z}{Q-2\int_0^{y_\mathrm{fj}^{}}\int_0^Hu_\mathrm{ax}(y,z)\,\mathrm{d}y\mathrm{d}z}\,,
\end{equation}
where $H$ is the height of the microchannel and $u_\mathrm{ax}(y,z)$ is the axial velocity field considered constant for the entire channel length $L$.

Similarly, one can estimate where the particle are separated into the outlet and center streams, by considering the outlet flowrate ratio
\begin{equation}
q_\mathrm{out}^{}=\frac{Q_{\mathrm{s,out}}^{}}{Q_{\mathrm{c,out}}^{}}\,,
\end{equation}
or in terms of the position of the streamline separating the side and the center outlet streams $y_\mathrm{fs}^{}$ (``fs'' stands for flow-splitting)
\begin{equation}
q_\mathrm{out}^{}(y_\mathrm{fs}^{})=\frac{2\int_0^{y_\mathrm{fs}^{}}\int_0^H u_\mathrm{ax}(y,z)\,\mathrm{d}y\mathrm{d}z}{Q-2\int_0^{y_\mathrm{fs}^{}}\int_0^Hu_\mathrm{ax}(y,z)\,\mathrm{d}y\mathrm{d}z}\,.
\end{equation}
resulting thus
\begin{equation}
y_\mathrm{sep}^{}=y_\mathrm{fs}^{}.
\end{equation}
The axial flow $u_\mathrm{ax}^{}(y,z)$ can be computed by considering the Poisson problem for the microchannel cross-section
\begin{equation}\label{eq:uax}
\bm\nabla_\perp^2 u_\mathrm{ax}^{}=\mathrm{const}\,,\qquad (y,z)\in[0,W]\times[0,H]
\end{equation}
where \mbox{$\bm{\nabla}_\perp^{}=[\partial_y^{},\partial_z^{}]^T_{}$}, and the constant such that the normalization condition
\begin{equation}\label{eq:uaxnorm}
\int_0^H\int_0^W u_\mathrm{ax}^{}(y,z)\,\mathrm{d}y\mathrm{d}z=Q\,,
\end{equation}
is verified.

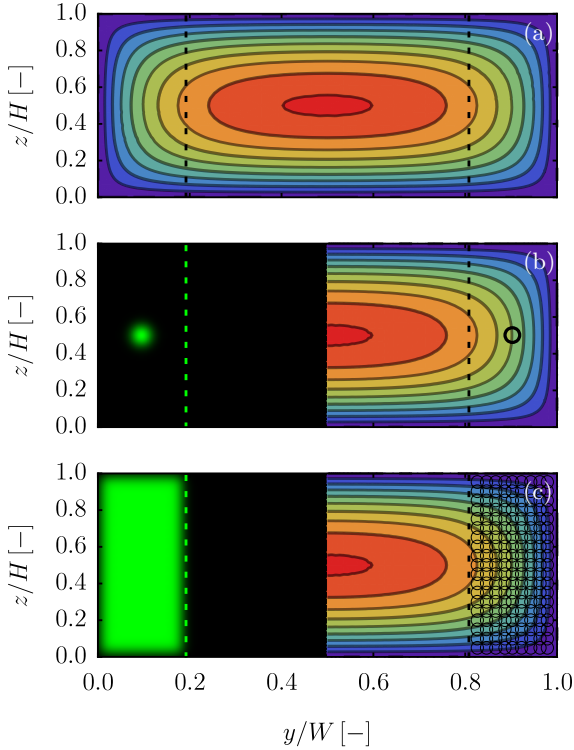
\begin{figure}[!!t]
%\fbox{\begin{picture}(226,148)
%\fbox{
\begin{picture}(226,280)
\put(-40,147){\includegraphics[width=10cm]{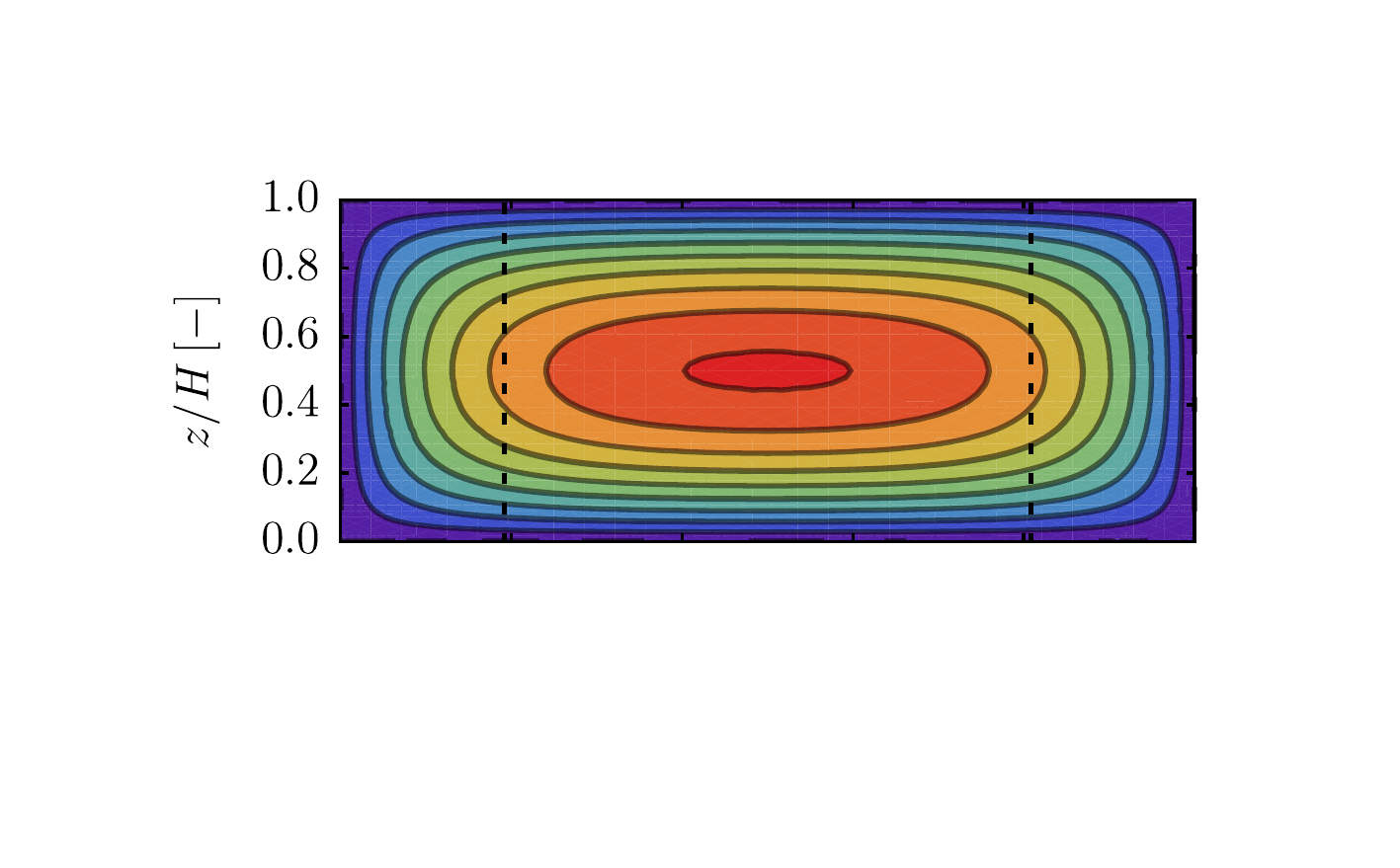}}
\put(-40,60){\includegraphics[width=10cm]{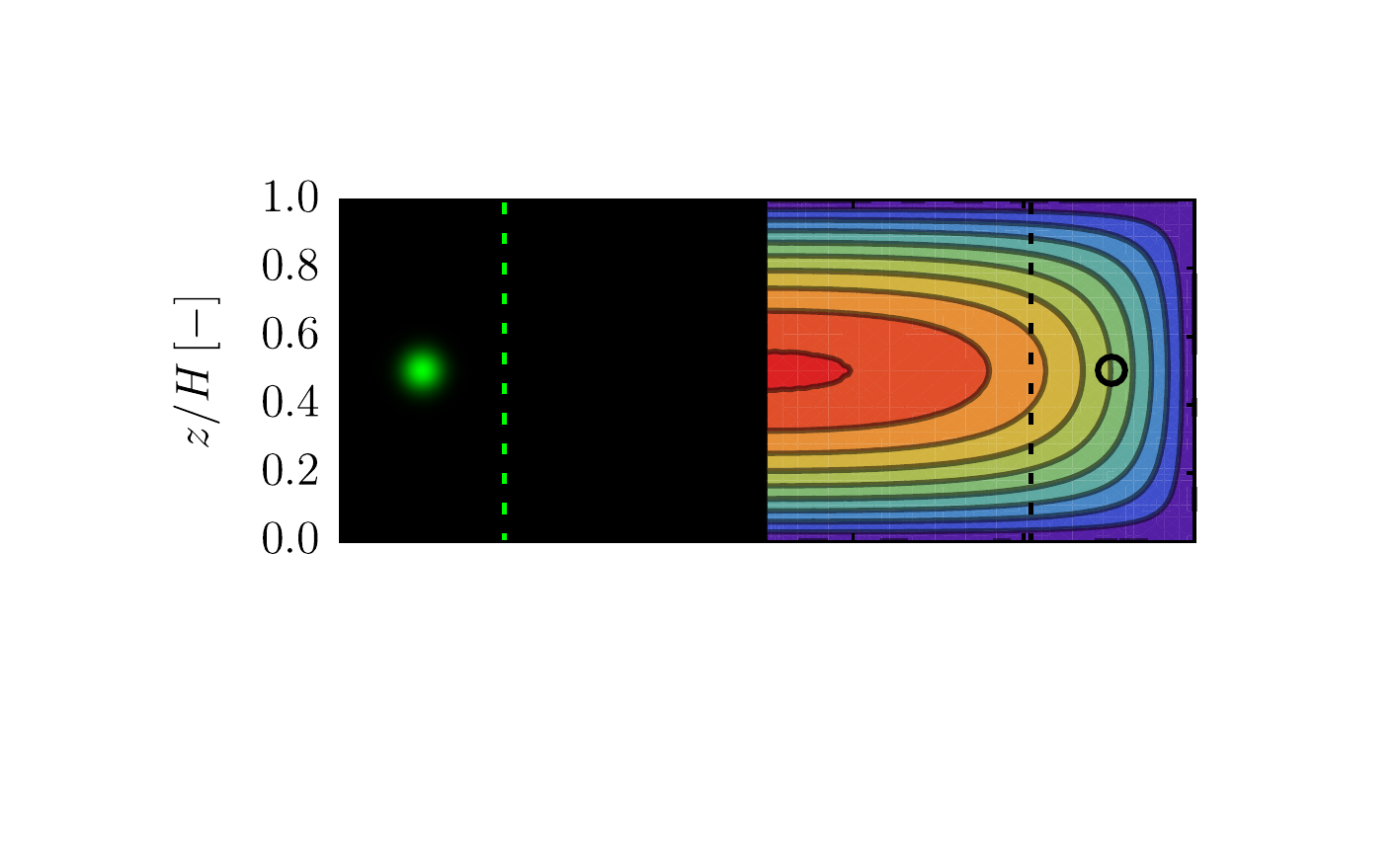}}
\put(-40,-27){\includegraphics[width=10cm]{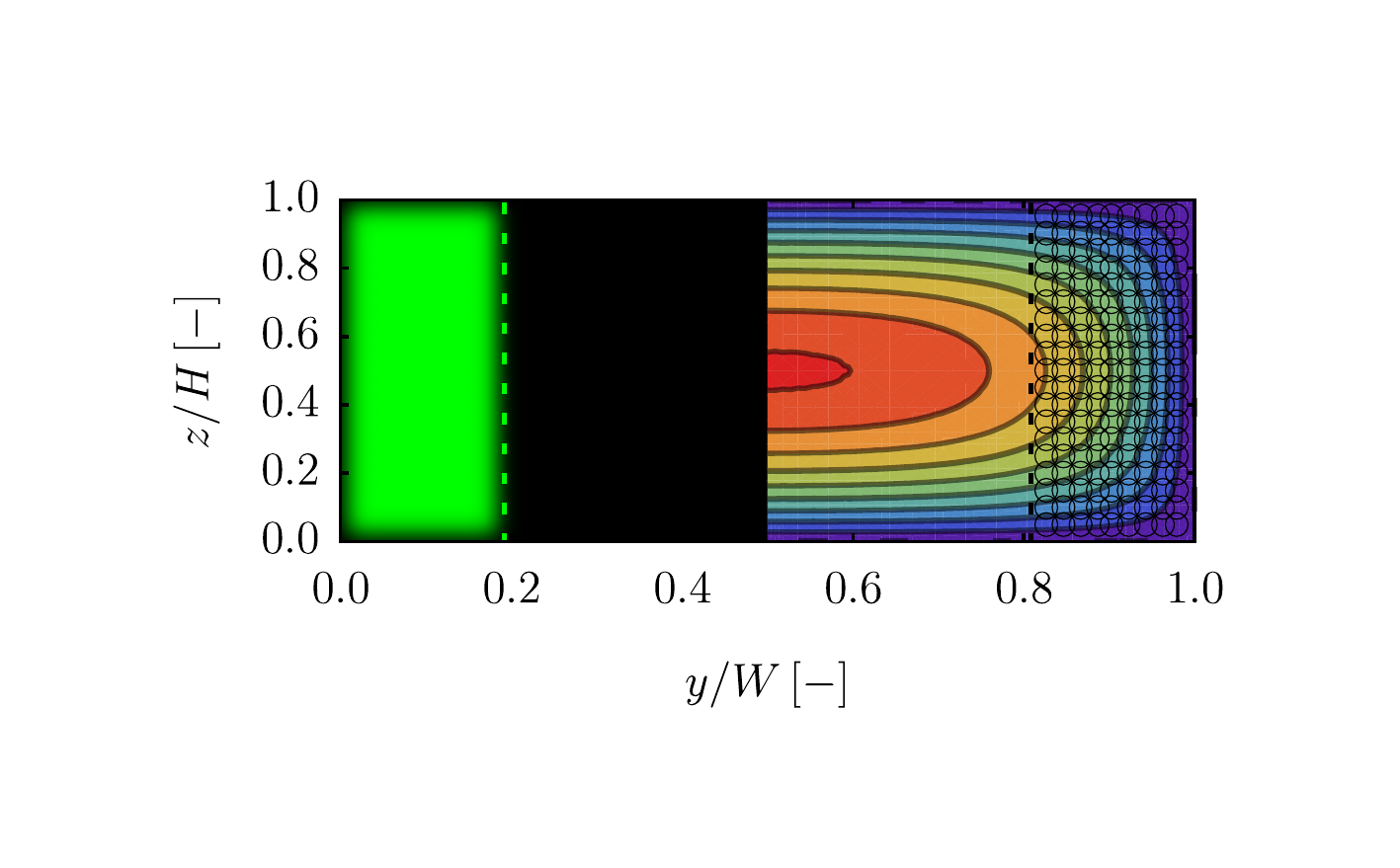}}
\put(190,273){\begin{color}{white}(a)\end{color}}
\put(190,99){\begin{color}{white}(c)\end{color}}
\put(190,186){\begin{color}{white}(b)\end{color}}
\end{picture}
%}
\caption{(Color Online) Axial velocity profile \mbox{$u_\mathrm{ax}^{}$} for a rectangular cross section with \mbox{$W=375\,\si{\micro m}$}, \mbox{$H=150\,\si{\micro m}$}, and \mbox{$Q=400\,\si{\micro l\,min^{-1}_{}}$} as function of the cross-section dimensionless position (a). Gaussian starting positions and corresponding distribution for prefocused particle streams (b) and non-prefocused particle streams (c). Vertical dotted lines are in the correspondence of $y_\mathrm{fj}^{}$.}
\label{fig:axialflow}
\end{figure}

Figure~\ref{fig:axialflow}(a) shows the axial velocity profile obtained by solving Eq.~\eqref{eq:uax} with the constraint Eq.~\eqref{eq:uaxnorm} for the case \mbox{$W=375\,\si{\micro m}$}, \mbox{$H=150\,\si{\micro m}$}, and \mbox{$Q=400\,\si{\micro l\,min^{-1}_{}}$}. The vertical dotted lines as well as the inlet position are computed by setting \mbox{$q_\mathrm{in}^{}=1/3$} in Eq.~\eqref{eq:qinyfj}. These are typical dimensions, flowrate and flowrate ratios used in acoustophoretic separation experiments~\cite{urbansky2017rapid}.

\paragraph*{Separation Indicators.}
The usual procedure to quantify the performance during acoustophoretic separation experiments is to measure the fraction of particles in the side and the center outlets downstream the separation channel by varying the voltage applied to the transducer.
This can be done by collecting the samples and counting the particles with either a Coulter Counter or a FACS machines, here the same quantification method is adopted by developing further the analysis of the MCDGM method.

The fraction of particles belonging to the $h$-th population that flow into the side-stream can be computed by considering the spatial marginal along the width of the channel, i.e. $y$-direction,
\begin{equation}
\rho^y_h(y,t\,|\,\q{}{0},t_0^{})=\sum_{k\in\mathcal{K}_h^{}}w_{hk}\,\mathcal{N}[y\,|\,m^y_{hk}(t\,|\,\q{}{0},t_0^{}),s^{yy}_{hk}(t\,|\,\q{}{0},t_0^{})]\,,
\end{equation}
and defining the side-stream recovery, henceforth $\mathrm{SSR}$, for the $h$-th population as the associated cumulative (omitting conditionals)
\begin{equation}
\mathrm{SSR}_h=\sum_{k\in\mathcal{K}_h^{}} w_{hk}\,\mathrm{SSR}_{hk}^{}\,,
\end{equation}
in which the side-stream recovery for the $k$-th gaussian of the $h$-th population is given by
\begin{equation}
\mathrm{SSR}_{hk}^{}=\frac{1}{2}\,\mathrm{erfc}\left[
\frac{m^y_{hk}(t_\mathrm{sep}^{}\,|\,\q{}{0},t_0^{})-y^{}_\mathrm{sep}}{\sqrt{2\,s^{yy}_{hk}(t\,|\,\q{}{0},t_0^{})}}\right]\,,
\end{equation}
where $t_\mathrm{sep}^{}$ is the separation time, and $y_\mathrm{sep}^{}=y_\mathrm{fs}^{}$ is the separation abscissa. These two parameters are constants that we assume depending solely on the flow conditions. Additionally, the separation time depends on the channel length $L$, that for the simulations is $L=4.3\,\si{cm}$.

When the fraction of particle in the center stream is measured, in place of using the SSR the center stream recovery
\begin{equation}
\mathrm{CSR}=1-\mathrm{SSR}\,,
\end{equation}
can be used. For the present example we use exclusively the side-stream recovery.

\paragraph*{Model Equations.} For the ensemble simulations of free-flow acoustophoretic separation, we consider the three-dimensional model
\begin{subequations}\label{eq:3dmodel}
\begin{align}
\dot{X}_h^{}(t)&=\,u_\mathrm{ax}^{}[Y_h^{}(t),Z_h^{}(t)]\,,\\
\dot{Y}_h^{}(t)&=\,\epsilon\,\mu(R_h^{},\tilde\kappa_h^{},\tilde\rho_h^{},\eta_\mathrm{f}^{})\,\sin\left[\frac{2\pi\,Y_h^{}(t)}{W}\right]\,,\\
\dot{Z}_h^{}(t)&=\,-\mu_\mathrm{g}(R_h^{},\rho_h^{},\rho_\mathrm{f}^{},\eta_\mathrm{f}^{})\,,
\end{align}
\end{subequations}
that takes into account for the axial flow, acoustophoresis and gravity. The gravitational mobility is
\begin{equation}
\mu_\mathrm{g}^{}(r,\rho_{}^{},\rho_\mathrm{f},\eta_\mathrm{f}^{})=\frac{2\,g\,r^2}{9\,\eta_\mathrm{f}^{}}(\rho_{}^{}-\rho_\mathrm{f}^{})\,.
\end{equation}
Since the aim is to show how the separation performance depend on the (measured) voltage on the transducer, we adopt a model that is linear with the square-voltage for the energy density in the factor \mbox{$\epsilon=\pi E_\mathrm{ac}^{}/W$}
\begin{equation}
E_\mathrm{ac}^{}=\alpha\,V_\mathrm{pp}^2\,,
\end{equation}
where the factor $\alpha$ should depend on the experimental conditions such as fluid properties, temperature, and generally on the system features.
In the present paper the value is fixed $\alpha=8.364$.

Here we consider up to four different types of microparticles, $h=\mathrm{PS5},\,\mathrm{PS7},\,\mathrm{WBC},\,\mathrm{RBC}$.
The histograms for \mbox{$\mathrm{RBC}$} and \mbox{$\mathrm{WBC}$} are those shown in Sec.~\ref{sec:SIPmodel}, while \mbox{$\mathrm{PS5}$} and \mbox{$\mathrm{PS7}$} are polystyrene particle with diameters \mbox{$5\,\si{\micro m}$} (\mbox{$m^r_\mathrm{PS5}=2.5\,\si{\micro m}$}) and \mbox{$7\,\si{\micro m}$} (\mbox{$m^r_\mathrm{PS7}=3.5\,\si{\micro m}$}), respectively, with a standard deviation (assuming a single kernel) $\sigma^r_{h}=7\%\,m^r_h$, the compressibility \mbox{$\kappa=273\,\si{TPa^{-1}_{}}$} and the density \mbox{$\rho=1058\,\si{kg\,m^{-3}_{}}$} are given by~\cite{cushing2017ultrasound}.

\begin{figure*}[!!t]
%\fbox{\begin{picture}(226,148)
%\fbox{
\begin{picture}(500,413)
%\begin{picture}(226,413)
\put(-40,255){\includegraphics[width=10cm]{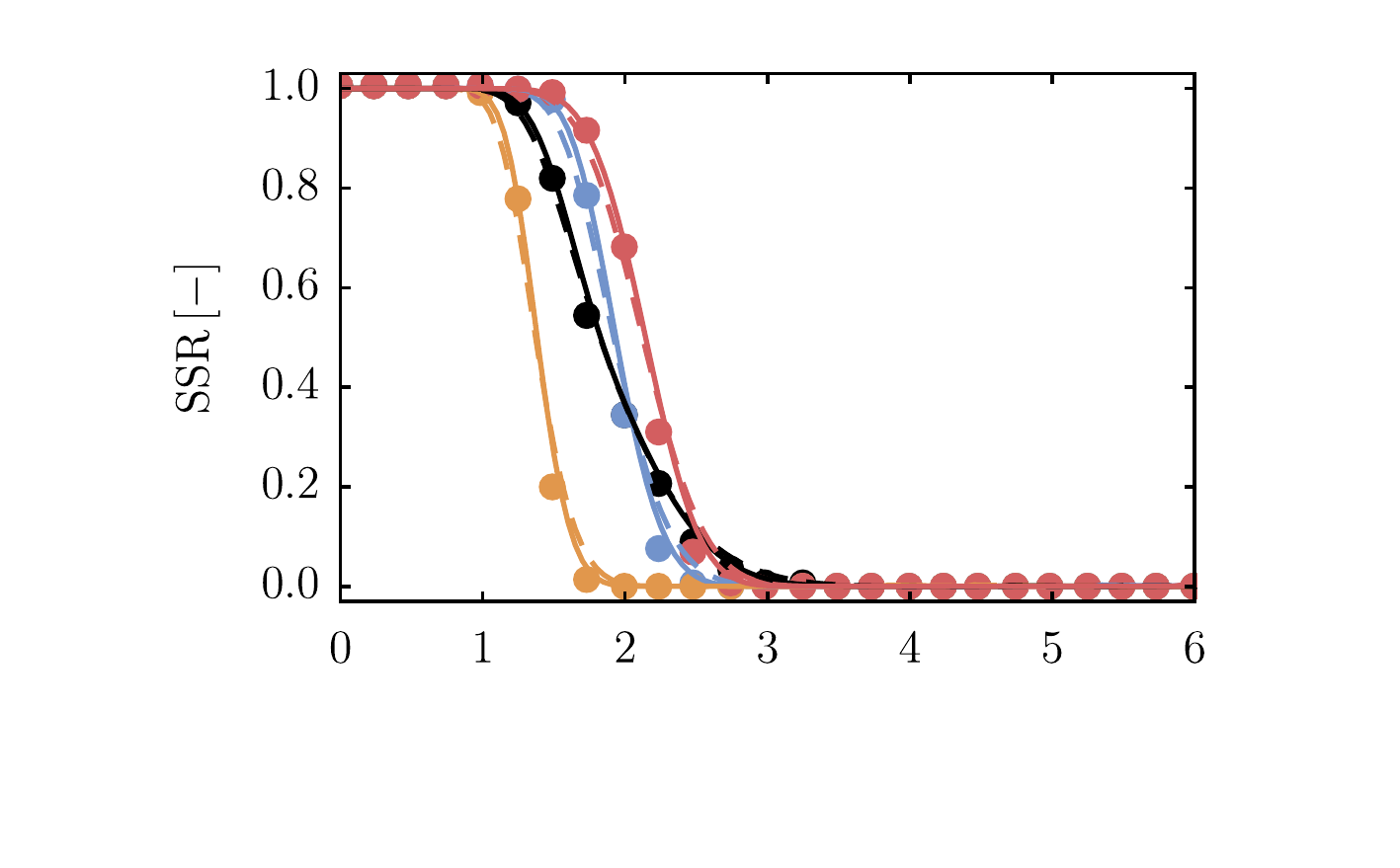}}
\put(-40,120){\includegraphics[width=10cm]{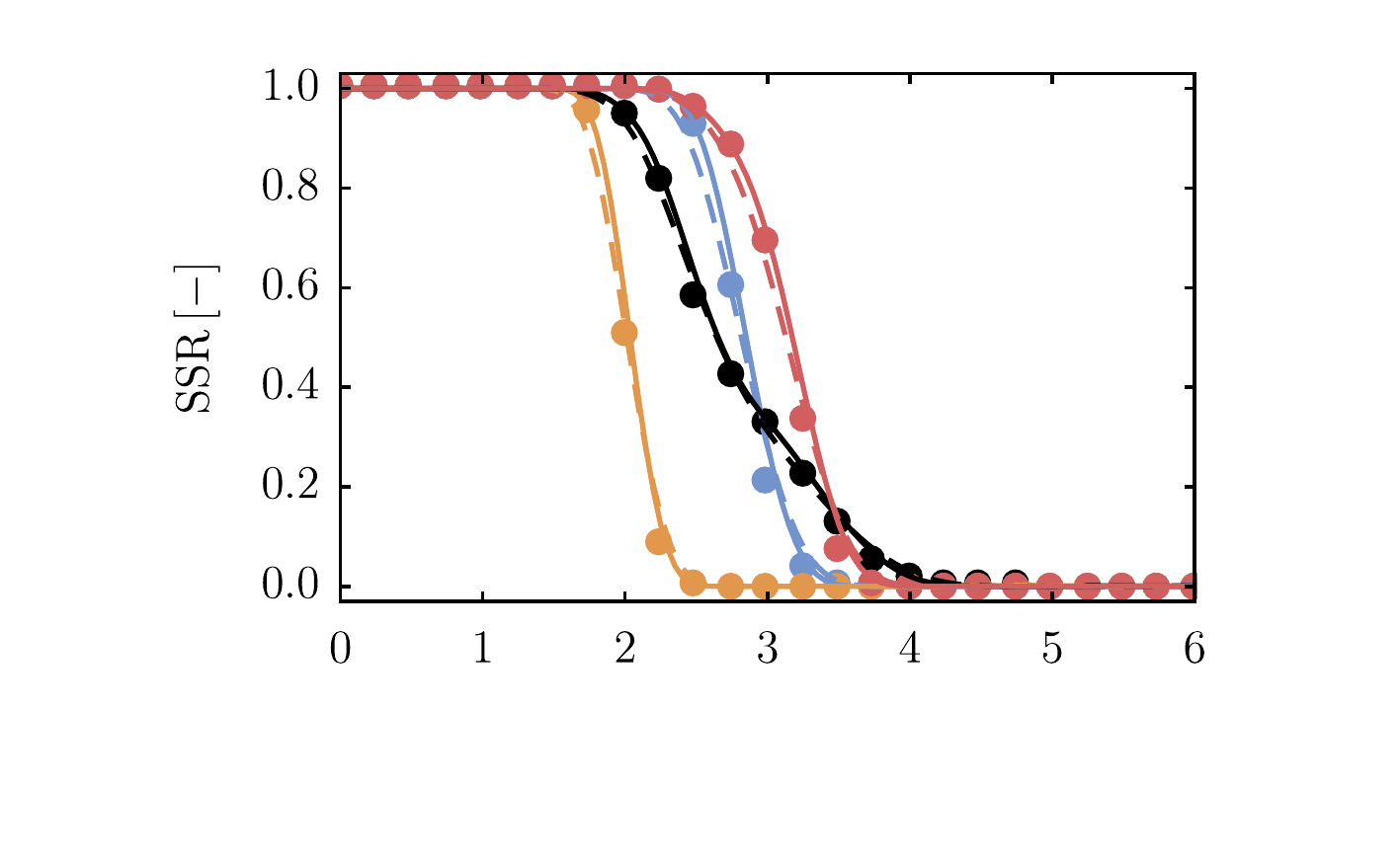}}
\put(-40,-15){\includegraphics[width=10cm]{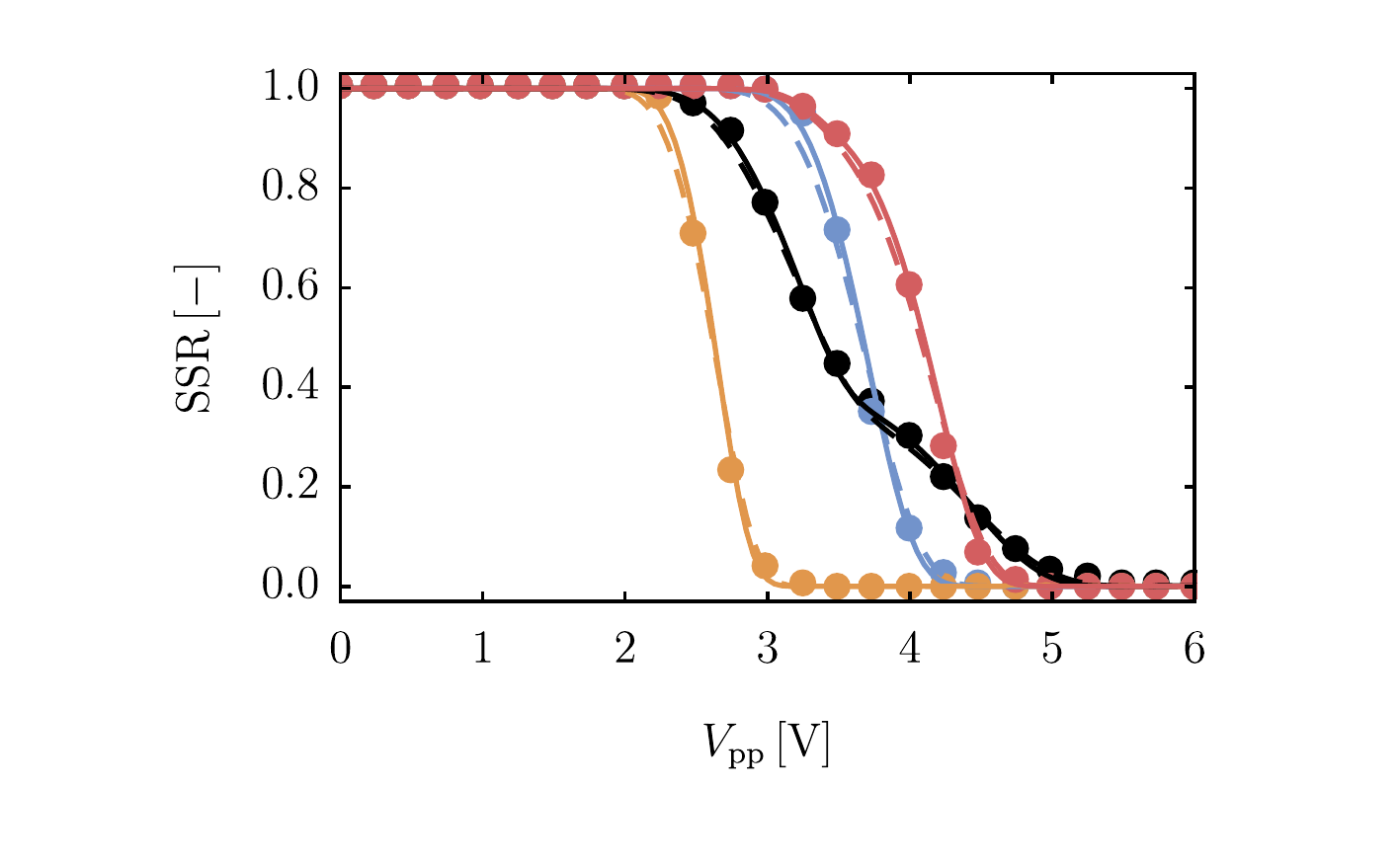}}
\put(230,255){\includegraphics[width=10cm]{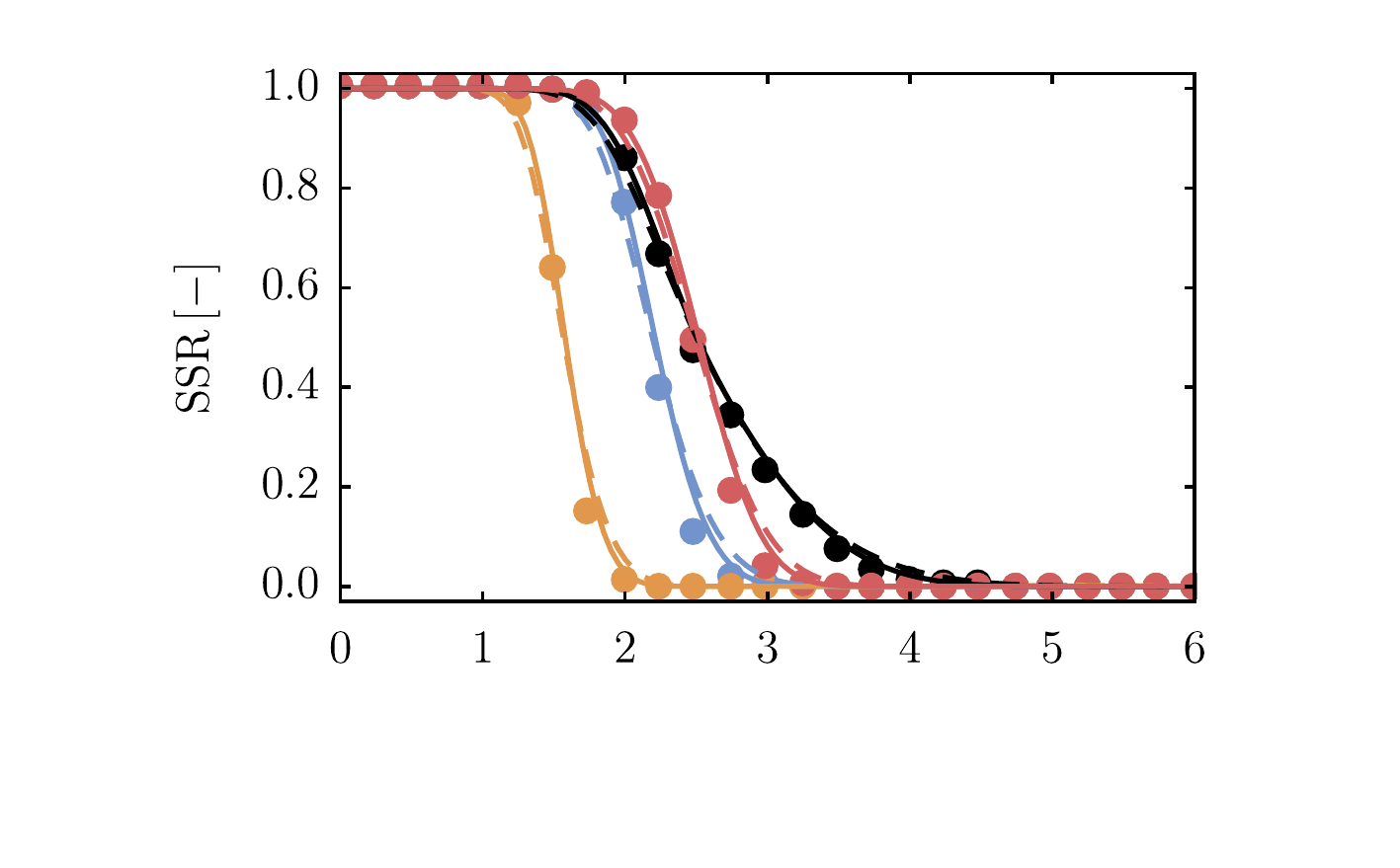}}
\put(230,120){\includegraphics[width=10cm]{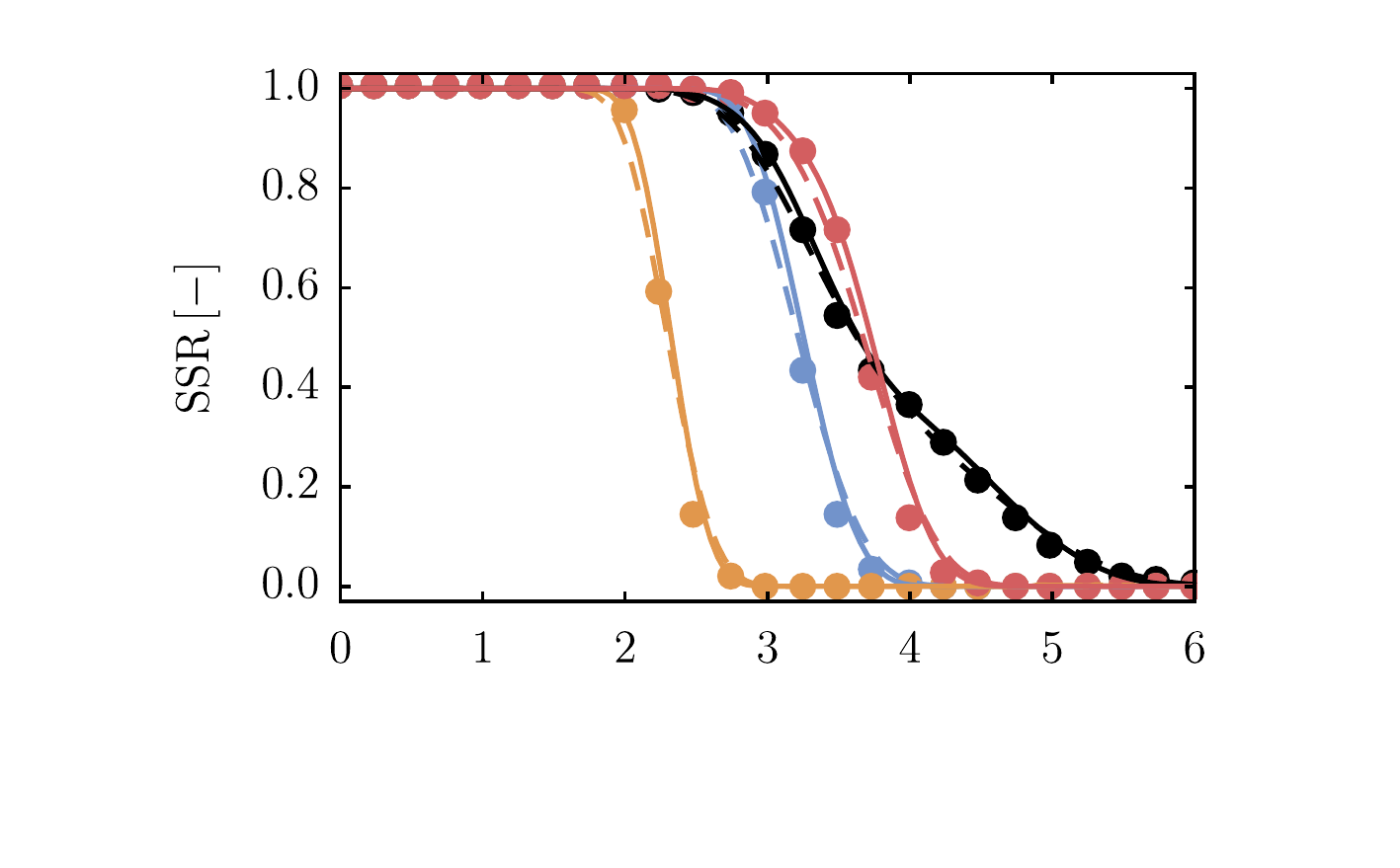}}
\put(230,-15){\includegraphics[width=10cm]{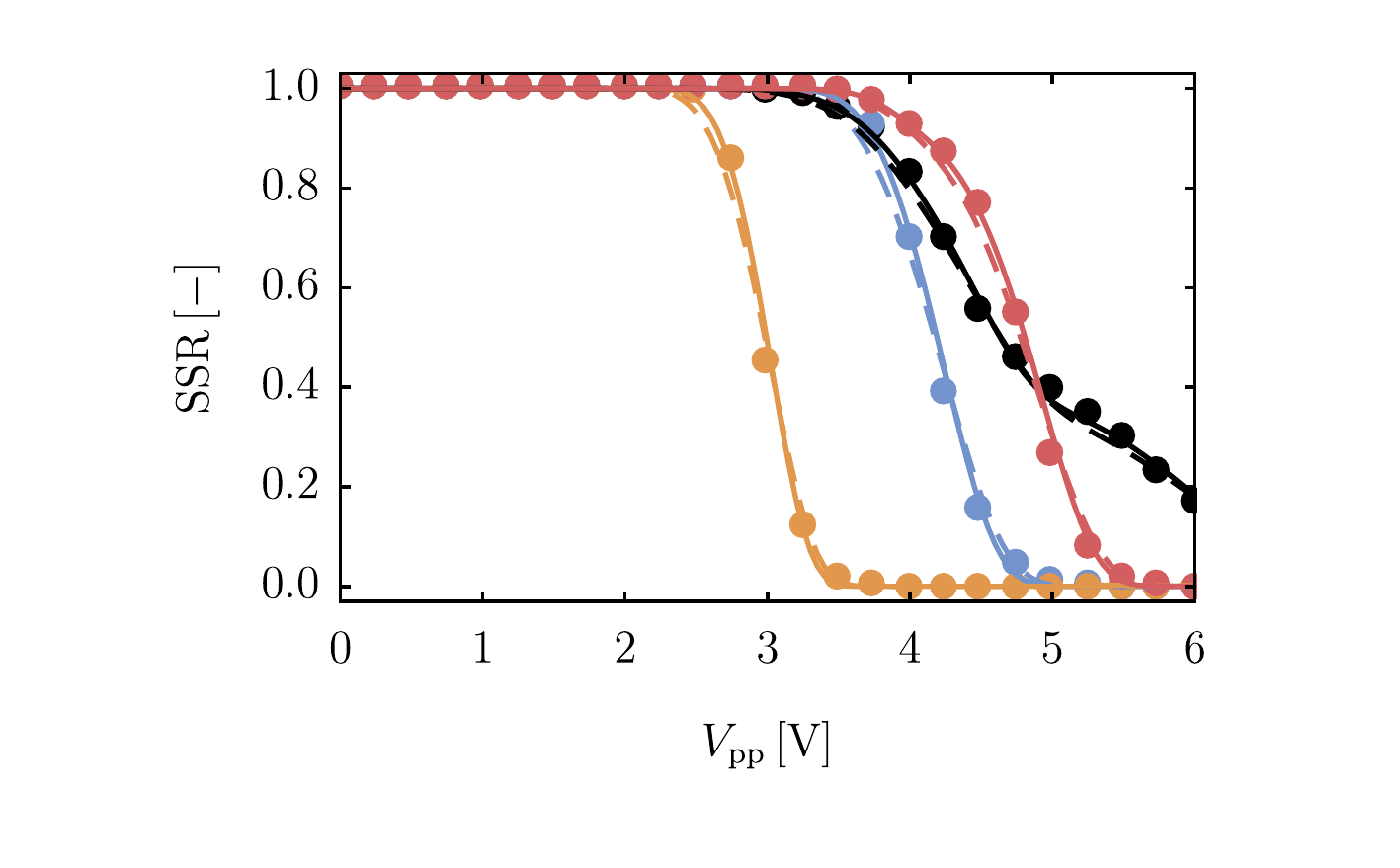}}
\put(158,405){\mbox{$q_\mathrm{out}^{}=1/3$}}
\put(158,270){\mbox{$q_\mathrm{out}^{}=1/1$}}
\put(158,135){\mbox{$q_\mathrm{out}^{}=3/1$}}
\put(428,405){\mbox{$q_\mathrm{out}^{}=1/3$}}
\put(428,270){\mbox{$q_\mathrm{out}^{}=1/1$}}
\put(428,135){\mbox{$q_\mathrm{out}^{}=3/1$}}
\end{picture}
%}
\caption{(Color Online) Side-stream recovery \mbox{$\mathrm{SSR}$} for prefocused streams as function of the voltage \mbox{$V_\mathrm{pp}^{}$} for \mbox{$q_\mathrm{in}^{}=1/3$}, \mbox{$\mathrm{SIP\%}=0\%$} (left) and \mbox{$\mathrm{SIP\%}=30\%$} (right), different values of \mbox{$q_\mathrm{out}^{}$}, and for different microparticles: \mbox{$\mathrm{PS}5\si{\micro m}$} beads (blue), \mbox{$\mathrm{PS}7\si{\micro m}$} beads (orange), \mbox{$\mathrm{WBC}$} (black), and \mbox{$\mathrm{RBC}$} (red). Symbols are computed from ensemble simulations, lines from the MCDGM method applied to Eqs.~\eqref{eq:WBCRBCmodel}-\eqref{eq:additionals} (dashed), or Eqs.~\eqref{eq:WBCRBCmodel} with correction Eq.~\eqref{eq:effvel} (continuous).
}
\label{fig:SSR}
\end{figure*}
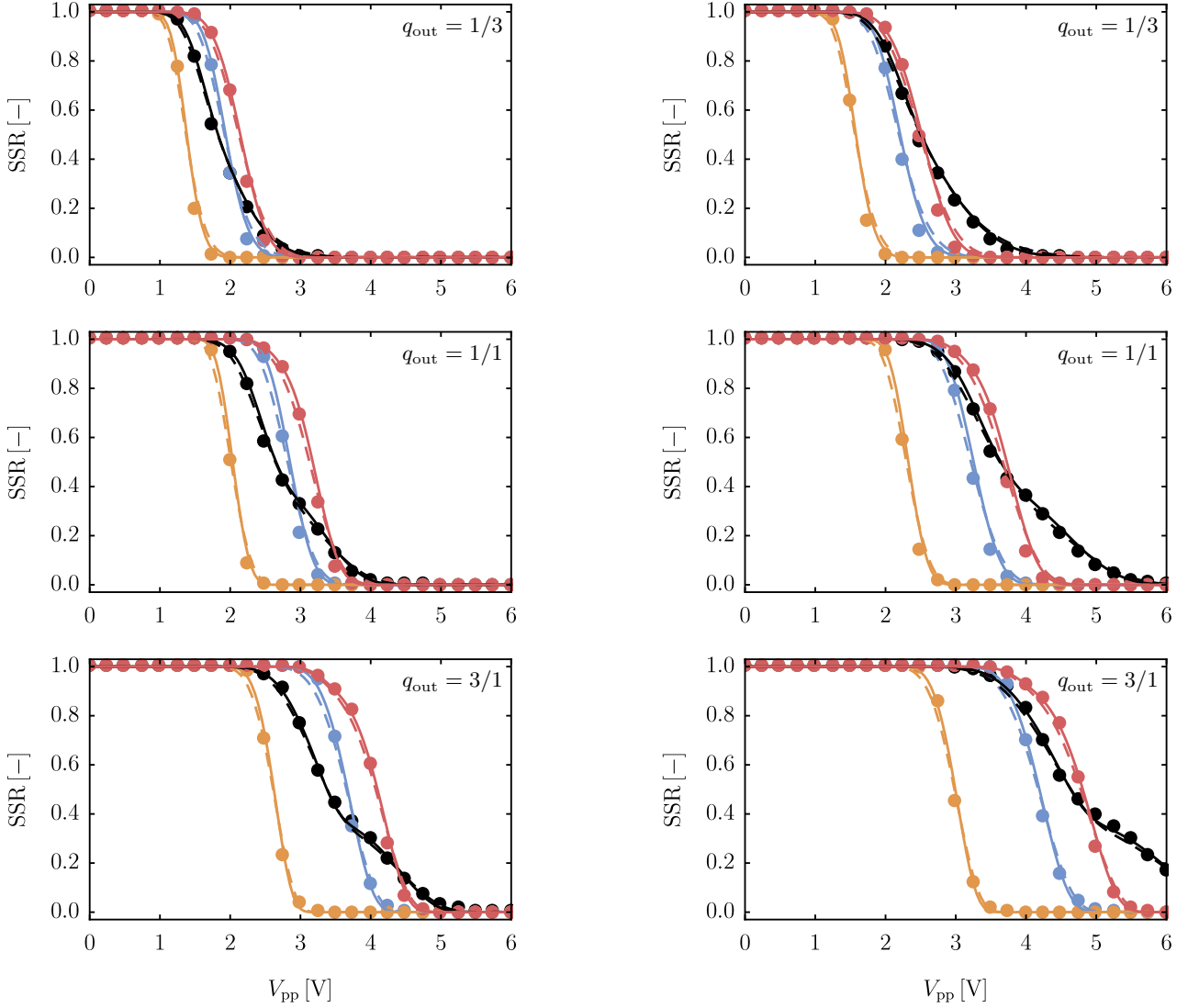

The MCDGM method applied to Eq.~\eqref{eq:3dmodel} yields cumbersome equations, so here we restrict the MCDGM analysis to two cases: (i)~a corrected plug-flow model for which \mbox{$t_\mathrm{sep}^{}=L/V$} where
\begin{equation}\label{eq:effvel}
V=\frac{1}{y_\mathrm{fs}^{}-y_\mathrm{in}^{}}\int_{y_\mathrm{in}^{}}^{y_\mathrm{fs}^{}}u(y',z_\mathrm{in}^{})\,\mathrm{d}y\,,
\end{equation}
is the average particle velocity between the inlet position $y_\mathrm{in}$ at height $z_\mathrm{in}^{}$ and the abscissa where the side and center outlet split,
and (ii)~a 2D model corresponding to disregard the equation for the $z$-component in Eq.~\eqref{eq:3dmodel}.
We expect that for either moderate/weak buoyant forces (as the overwhelming majority of the cases for polymer microbeads and cells) or fast passages in the separation channel, neglecting the vertical component in Eqs.~\eqref{eq:3dmodel} is a good approximation.
For the case (i) the MCDGM equations reduces to Eqs.~\eqref{eq:WBCRBCmodel}, while for the case (ii) there are additional equations to Eqs.~\eqref{eq:WBCRBCmodel}
\begin{subequations}\label{eq:additionals}
\begin{align}
\dot{m}^x_{hk}&=\,u_\mathrm{ax}^{}(m^y_{hk},z^0_{hk})\,\\
\dot{s}^{xx}_{hk}&=\,2\,\partial_y^{}u_\mathrm{ax}^{}(m^y_{hk},z^{0}_{hk})\,s^{xy}_{hk}\,,\\
\dot{s}^{xy}_{hk}&=\,\epsilon\,\mu_{hk}^{}\frac{2\pi}{W}\cos\left(\frac{2\pi\,m^y_{hk}}{W}\right)+\nonumber\\
&+\,\epsilon\,\mu'_{hk}\,\sin\left(\frac{2\pi\,m^y_{hk}}{W}\right)\,s^{xr}_{hk}+\nonumber\\
&+\,\partial_y^{}u_\mathrm{ax}^{}(m^y_{hk},z^{0}_{hk})\,s^{yy}_{hk}\,,\\
\dot{s}^{xr}_{hk}&=\,\partial_y^{}u_\mathrm{ax}^{}(m^y_{hk},z^{0}_{hk})\,s^{yr}_{hk}\,,
\end{align}
\end{subequations}
where $z^0_{hk}$ is the vertical position of the $k$-th kernel in the $h$-th population.

% \begin{figure}[!!t]
% %\fbox{\begin{picture}(226,148)
% %\fbox{
% \begin{picture}(226,413)
% \put(-40,255){\includegraphics[width=10cm]{separation_qout3_1_SIP030.pdf}}
% \put(-40,120){\includegraphics[width=10cm]{separation_qout1_1_SIP030.pdf}}
% \put(-40,-15){\includegraphics[width=10cm]{separation_qout1_3_SIP030.pdf}}
% \put(158,405){\mbox{$q_\mathrm{out}^{}=3/1$}}
% \put(158,270){\mbox{$q_\mathrm{out}^{}=1/1$}}
% \put(158,135){\mbox{$q_\mathrm{out}^{}=1/3$}}
% \end{picture}
% %}
% \caption{(Color Online) Side Stream Recovery \mbox{$\mathrm{SSR}$} as function of the voltage \mbox{$V_\mathrm{pp}^{}$} for \mbox{$q_\mathrm{in}^{}=3$}, \mbox{$\mathrm{SIP\%}=30\%$}, different values of \mbox{$q_\mathrm{out}^{}$}, and for different microparticles: \mbox{$\mathrm{PS}5\si{\micro m}$} beads (blue), \mbox{$\mathrm{PS}7\si{\micro m}$} beads (orange), \mbox{$\mathrm{WBC}$} (black), and \mbox{$\mathrm{RBC}$} (red). Symbols are computed from ensemble simulations, lines from the MCDGM method applied to Eqs.~\eqref{eq:WBCRBCmodel}-\eqref{eq:additionals} (dashed), or Eqs.~\eqref{eq:WBCRBCmodel} with correction Eq.~\eqref{eq:effvel} (continuous).
% }
% \label{fig:SSRSIP030}
% \end{figure}

\paragraph*{Prefocusing Strategy.}
For prefocused streams at the inlet section of the separation channel since it is expected that the particles focus at one quarter of the channel width and at half-height, it can assume
\begin{equation}
m^y_{0,hk}=\frac{1}{2}\,y_\mathrm{fj}^{}\,,\qquad m^z_{0,hk}=\frac{1}{2}H\,.
\end{equation}
The initial position in the $y$-th direction is computed by the calculation illustrated in the previous paragraph for \mbox{$q_\mathrm{in}^{}=1/3$}, and it is \mbox{$y_\mathrm{fj}^{}\simeq 0.192\,W$}.
The initial conditions for the particle ensemble simulations are
\begin{subequations}
\begin{align}
X_{hk}^{}(0)&\sim\,\delta(\cdot)\,,\\
Y_{hk}^{}(0)&\sim\,\mathcal{N}(\:\cdot \:;m^y_{0,hk},s^{yy}_{0,hk})\,,\\
Z_{hk}^{}(0)&\sim\,\mathcal{N}(\:\cdot \:;m^z_{0,hk},s^{zz}_{0,hk})\,,
\end{align}
\end{subequations}
where $\delta$ is the Dirac-delta distribution, \mbox{$m^y_{0,hk}$} and \mbox{$m^z_{0,hk}$} are the average positions given above, and \mbox{$s^{zz}_{hk}=s^{yy}_{hk}=(\sigma^y_{hk})^2_{}$} with \mbox{$\sigma^y_{hk}=y_\mathrm{fj}^{}/12$} the variances.
The initial distribution is shown in Fig.~\ref{fig:axialflow}(b).

Figures~\ref{fig:SSR} show the side-stream recovery as function of the applied voltage when \mbox{$q_\mathrm{in}^{}=1/3$}, two SIP concentrations and three cases of \mbox{$q_\mathrm{out}^{}$} for four different types of microparticles, two polymer microbeads PS5 and PS7 and two cells WBC and RBC.
The particle ensemble simulations are indicated with the symbols, the MCDGM method applied to the 1D model is indicated by the solid lines, while it is indicated with dashed lines for the 2D model.
In all the cases both the 1D and 2D models can approximate the numerical data quite well, showing the approximation properties of the MCDGM method, its robustness in terms of the parameter variations, and the validity of the effective velocity assumption Eq.~\eqref{eq:effvel}.

\paragraph*{No-Prefocusing Strategy.}
For non-prefocused particle streams the particle ensemble simulations are initializated by the conditions
\begin{subequations}
\begin{align}
X_{hk}^{}(0)&\sim\,\delta(\:\cdot\:)\,,\\
Y_{hk}^{}(0)&\sim\,\mathcal{U}(\:\cdot\:;m^r_{hk},y_\mathrm{fs}^{}-m^r_{hk})\,,\\
Z_{hk}^{}(0)&\sim\,\mathcal{U}(\:\cdot\:;m^r_{hk},H-m^r_{hk})\,,
\end{align}
\end{subequations}
where \mbox{$\mathcal{U}(\:\cdot\:;a,b)$} is a uniform distribution between $a$ and $b$. Note that a small portion of the cross-section has been excluded from the particle distribution, and this corresponds to the fact that the particles cannot have a distant from the walls less than the average radius.
For that regards the starting position of the gaussians chosen a resolution $n_y^{}$ in the $y$-direction, the resolution in the $z$-direction is \mbox{$n_z^{}=\lfloor n_y^{}H/W\rfloor$}, it has for the average positions
\begin{subequations}
\begin{align}
m^{y}_{hk}&=\,h\,\Delta y\,,\qquad h=1\,,...\,,n_y-1^{}\\
m^{z}_{hk}&=\,k\,\Delta z\,,\qquad k=1\,,...\,,n_z-1^{}
\end{align}
\end{subequations}
where \mbox{$\Delta y=y_\mathrm{fj}^{}/n_y^{}$} and \mbox{$\Delta z=H/n_z^{}$}, while for the variances
\begin{equation}
\sigma^{y}_{hk}=\Delta y\,\log 2\,,\qquad \sigma^{z}_{hk}=\Delta z\,\log 2\,,
\end{equation}
where \mbox{$\log 2$} is a factor chosen as to accommodate for the smootheness of the spatial distribution.
The initial distribution is shown in Fig.~\ref{fig:axialflow}(c).

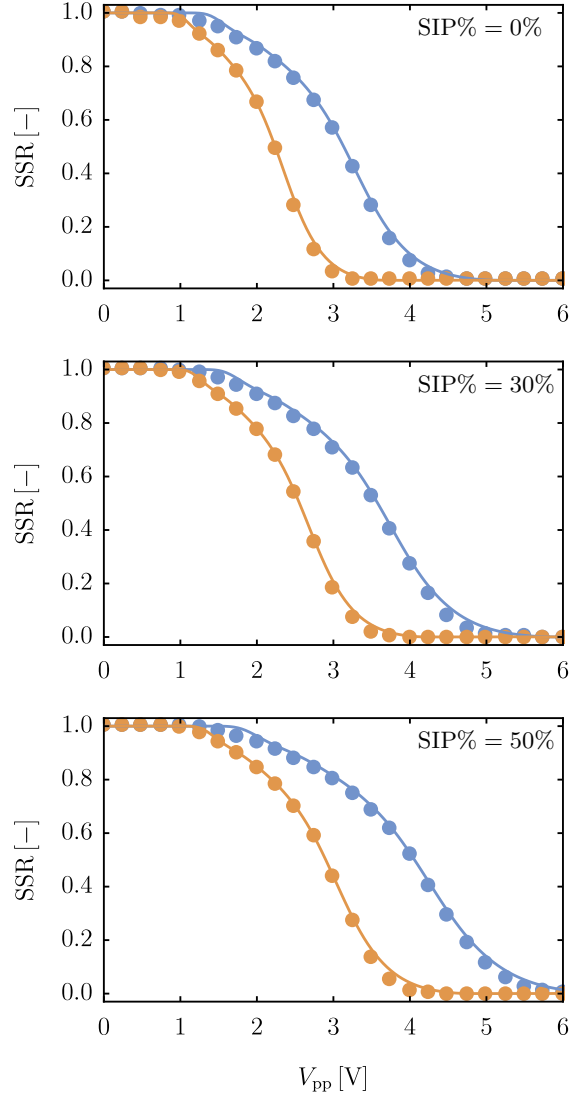
\begin{figure}[!!t]
%\fbox{\begin{picture}(226,148)
%\fbox{
\begin{picture}(226,413)
\put(-40,255){\includegraphics[width=10cm]{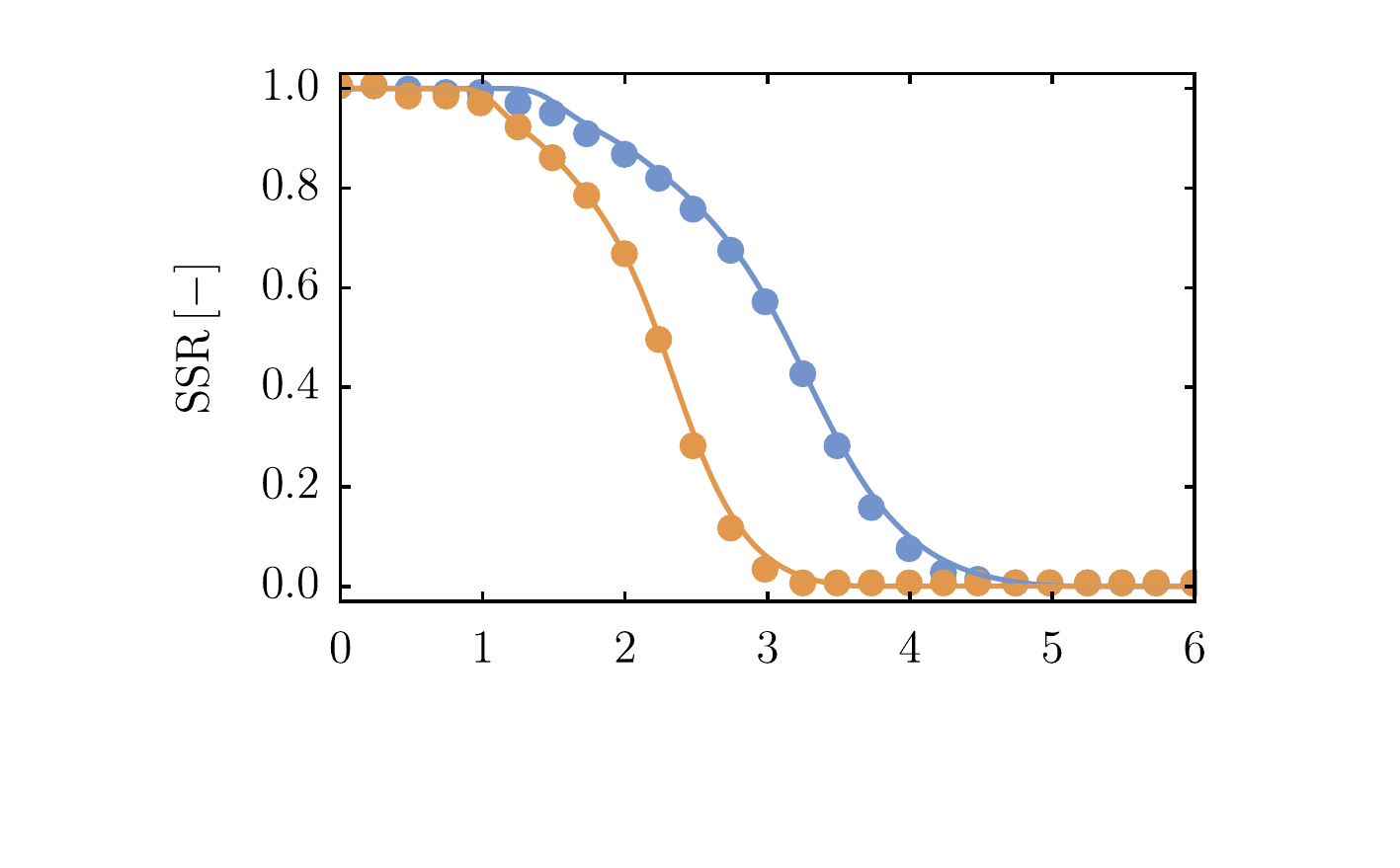}}
\put(-40,120){\includegraphics[width=10cm]{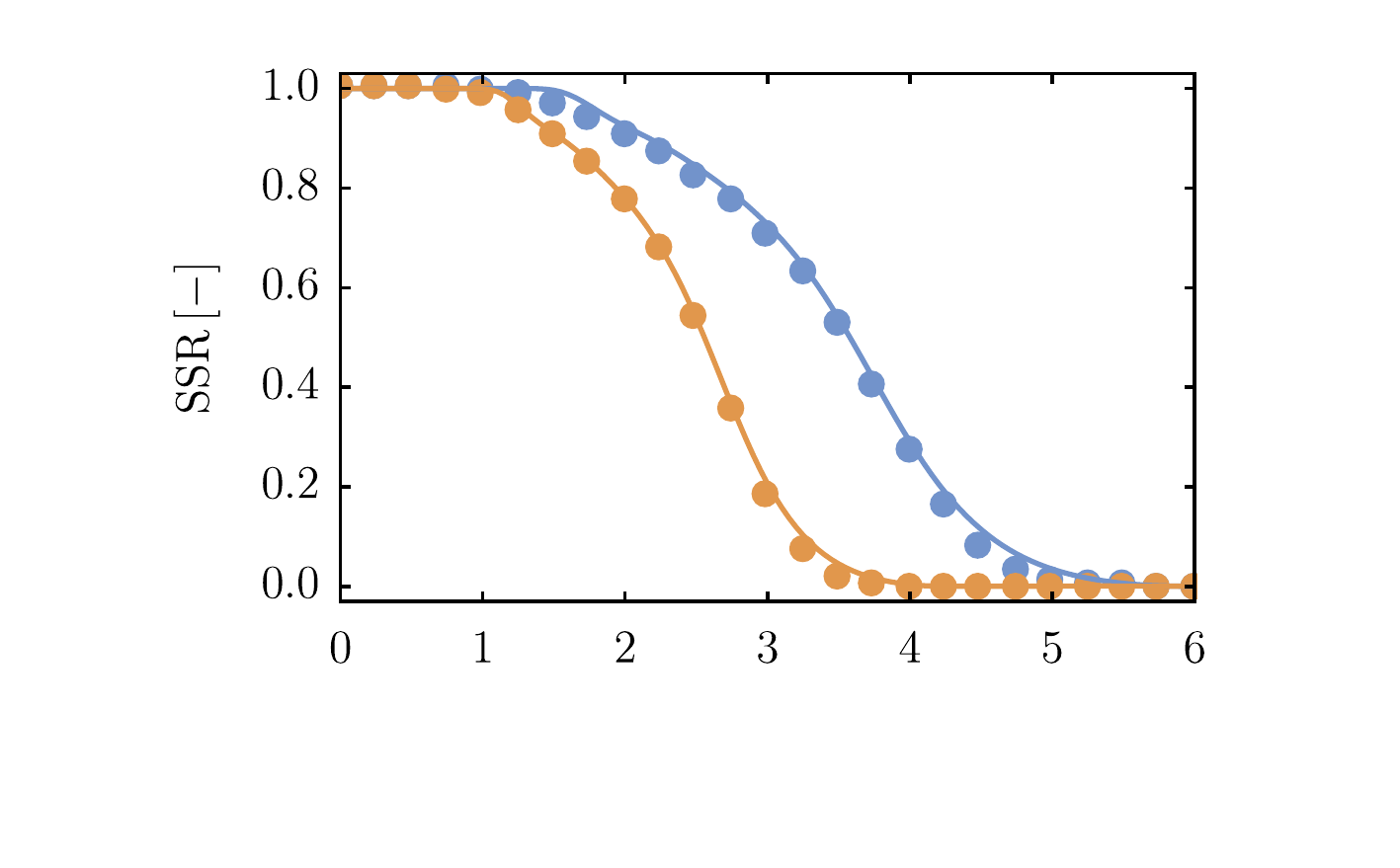}}
\put(-40,-15){\includegraphics[width=10cm]{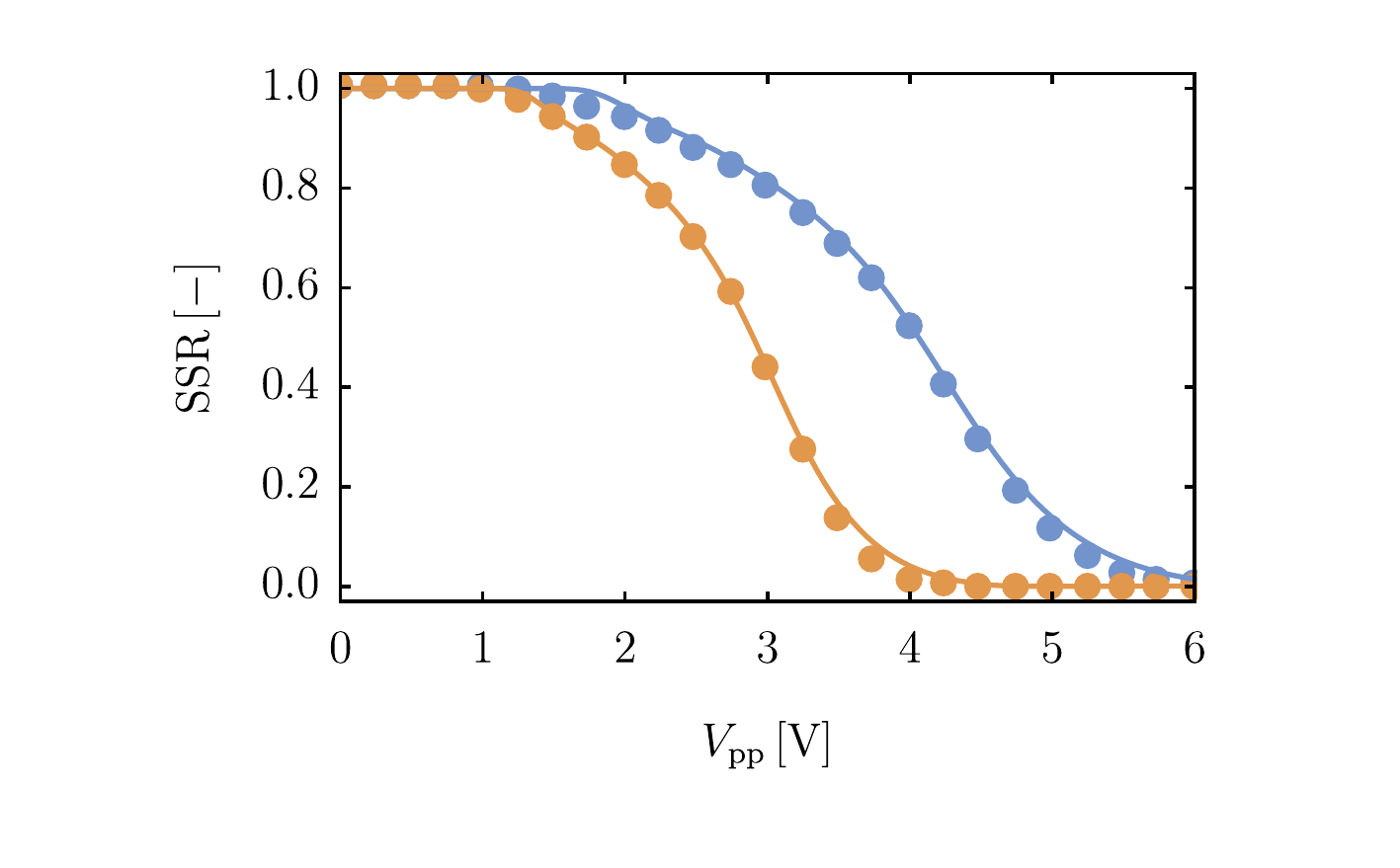}}
\put(148,405){\mbox{$\mathrm{SIP}\%=0\%$}}
\put(148,270){\mbox{$\mathrm{SIP}\%=30\%$}}
\put(148,135){\mbox{$\mathrm{SIP}\%=50\%$}}
\end{picture}
%}
\caption{(Color Online) Side-stream recovery \mbox{$\mathrm{SSR}$} for non-prefocused streams as function of the voltage \mbox{$V_\mathrm{pp}^{}$} for \mbox{$q_\mathrm{in}^{}=1/3$}, \mbox{$q_\mathrm{out}^{}=3/1$}, at different \mbox{$\mathrm{SIP\%}$} : \mbox{$\mathrm{PS}5\si{\micro m}$} beads (blue), \mbox{$\mathrm{PS}7\si{\micro m}$} beads (orange), \mbox{$\mathrm{WBC}$} (black), and \mbox{$\mathrm{RBC}$} (red). Symbols correspond to the particle ensemble simulations, lines correspond to the MCDGM method applied to Eqs.~\eqref{eq:WBCRBCmodel} with the correction Eq.~\eqref{eq:effvel}.
}
\label{fig:SSR_noprefoc}
\end{figure}

Figures~\ref{fig:SSR_noprefoc} shows the side-stream recovery as function of the applied voltage when \mbox{$q_\mathrm{in}^{}=1/3$}
for non-prefocused particle streams for two different polymer microbeads PS5 and PS7.
The particle ensemble simulations are indicated with the symbols, the MCDGM method applied to the 1D model is indicated by the solid lines.
Also for the case of non-prefocused particle streams the MCDGM method in both the 1D and 2D versions can approximate the numerical data quite well.
This confirms again the relialbility of the MCDGM method when applied to free-flow acoustophoretic separations.

\begin{figure*}[!!t]
%\fbox{\begin{picture}(226,148)
%\fbox{
\begin{picture}(500,413)
%\begin{picture}(226,413)
\put(-40,255){\includegraphics[width=10cm]{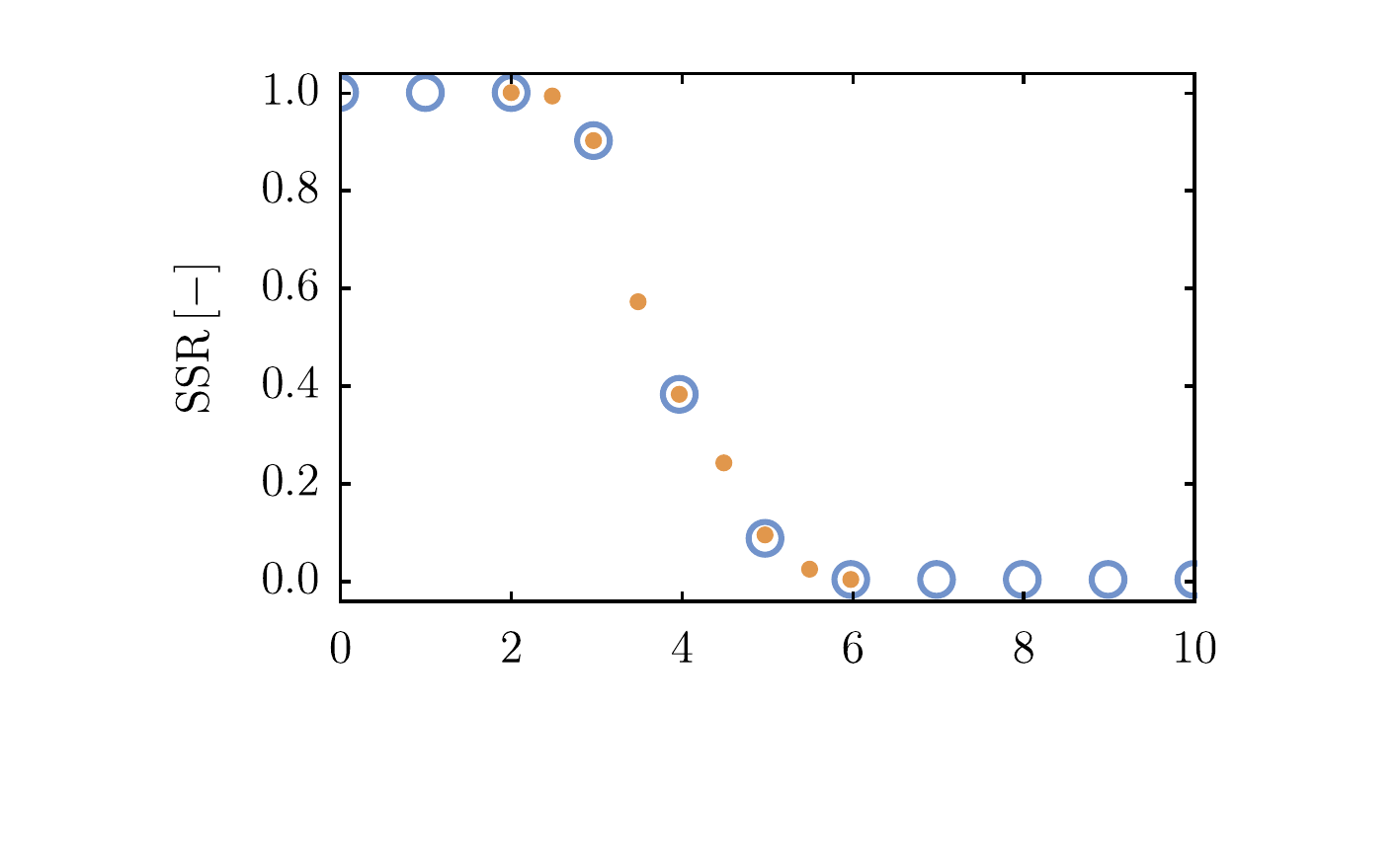}}
\put(-40,120){\includegraphics[width=10cm]{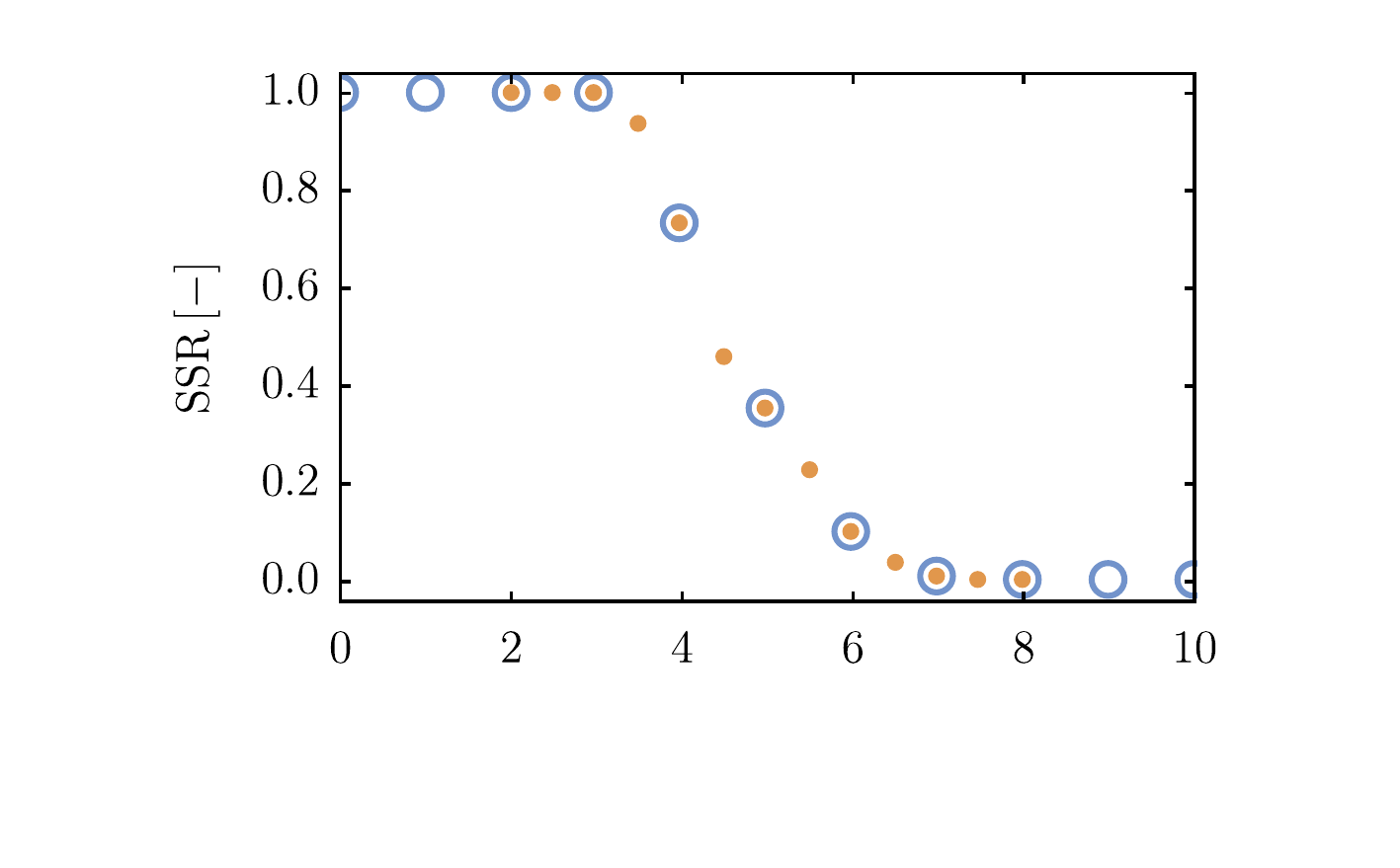}}
\put(-40,-15){\includegraphics[width=10cm]{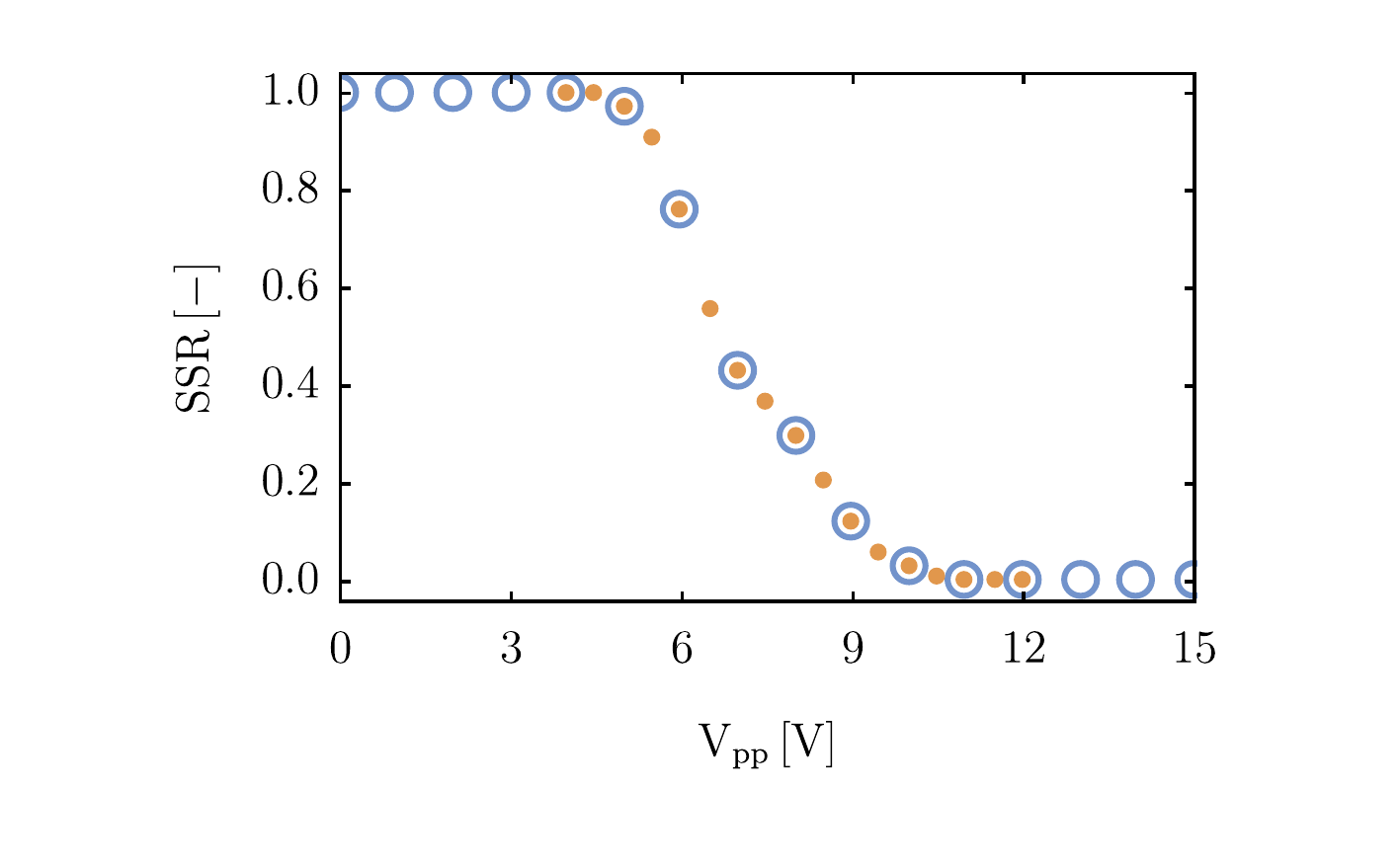}}
\put(230,255){\includegraphics[width=10cm]{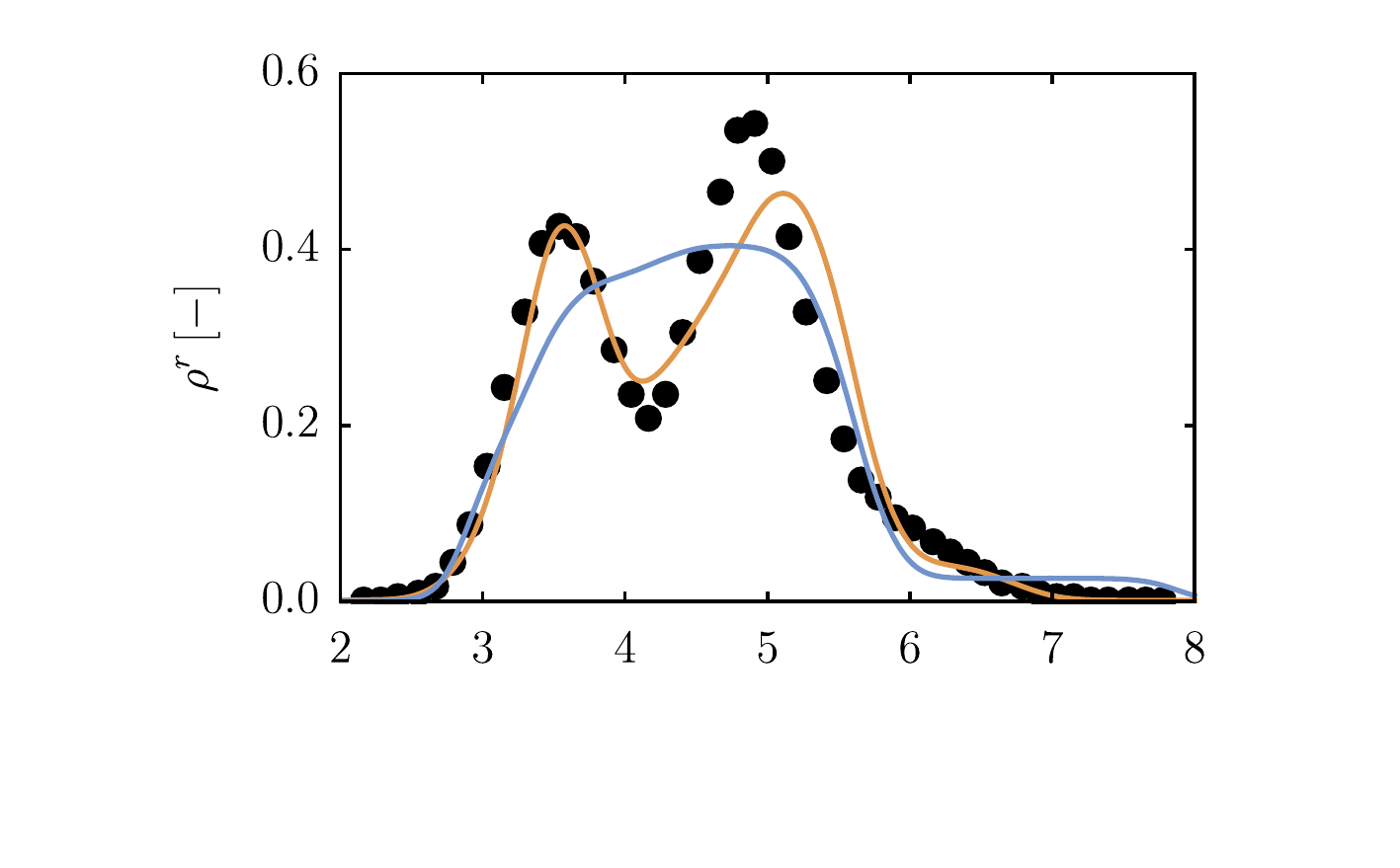}}
\put(230,120){\includegraphics[width=10cm]{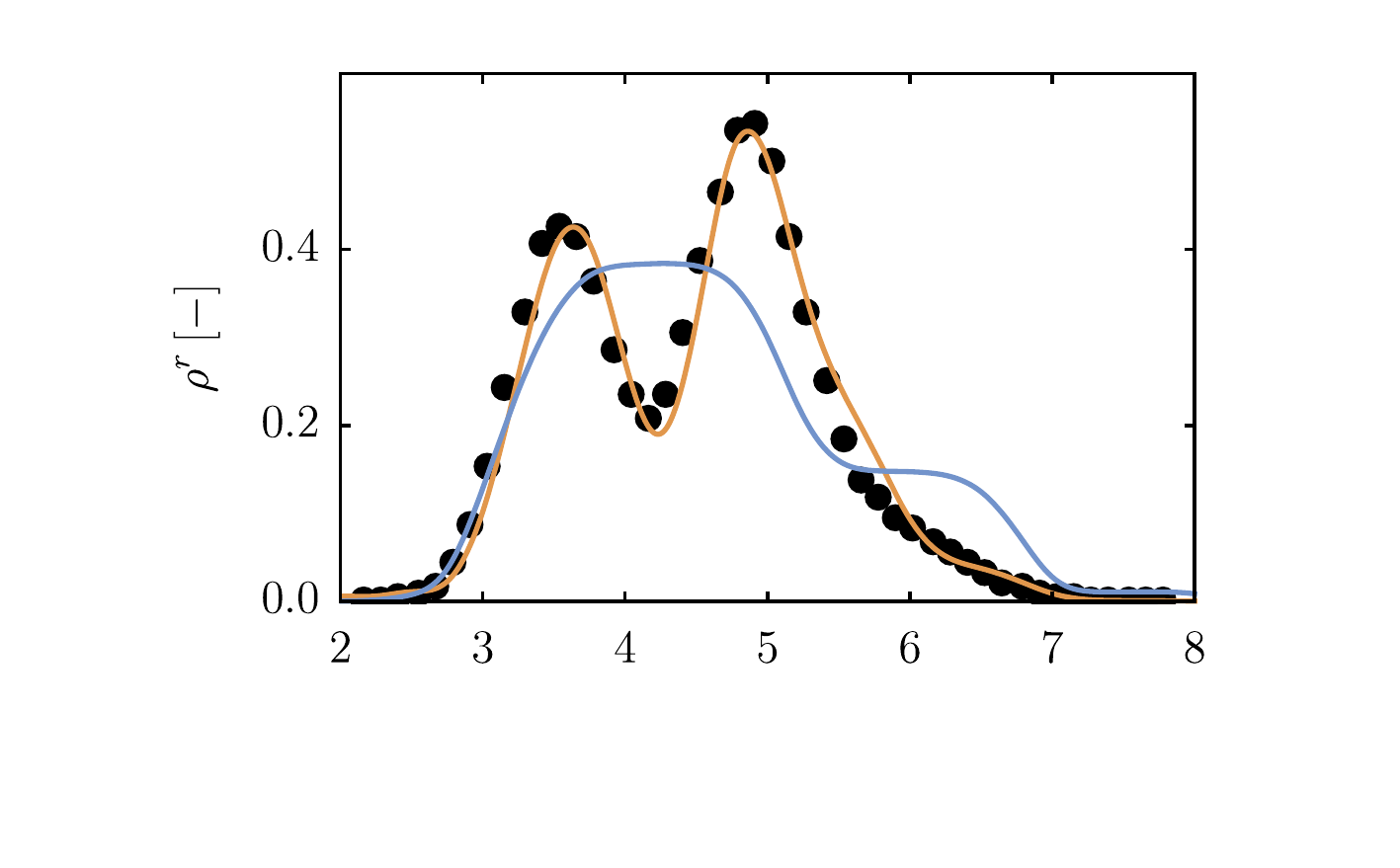}}
\put(230,-15){\includegraphics[width=10cm]{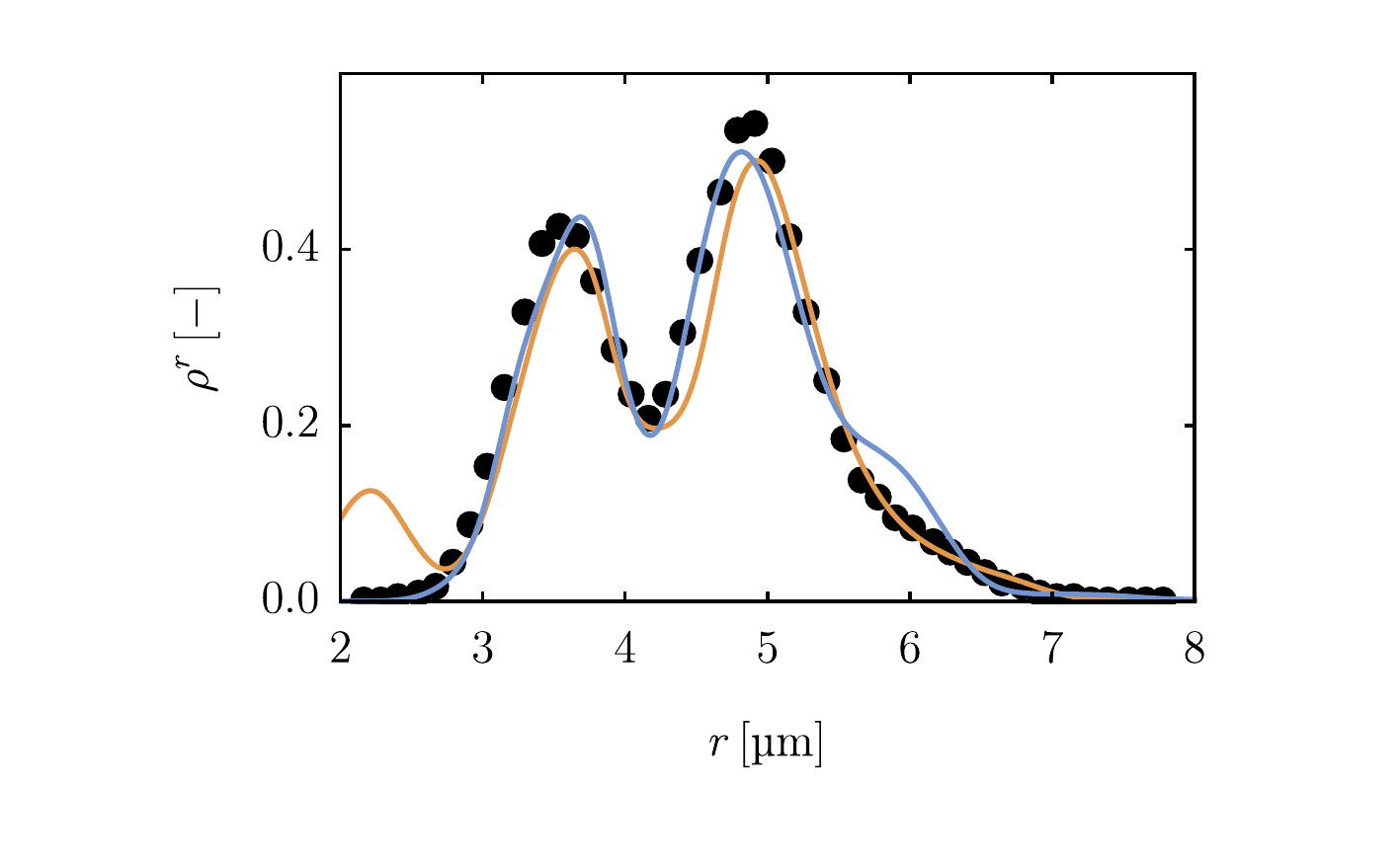}}
\put(150,405){\mbox{$\mathrm{SIP\%}=0\%$}}
\put(150,270){\mbox{$\mathrm{SIP\%}=25\%$}}
\put(150,135){\mbox{$\mathrm{SIP\%}=50\%$}}
\put(420,405){\mbox{$\mathrm{SIP\%}=0\%$}}
\put(420,270){\mbox{$\mathrm{SIP\%}=25\%$}}
\put(420,135){\mbox{$\mathrm{SIP\%}=50\%$}}
\end{picture}
%}
\caption{(Color Online) Size-histogram estimation (right) from SSR measurements as function of the applied voltage (left) for the WBC population.
Panels in the left column are the SSR measurements at different \mbox{$\mathrm{SIP}\%$}: blue cirlces indicate the full range of measurements, while the orange bullets indicate the dynamic range of measurements. Right column are the histograms computed for the different \mbox{$\mathrm{SIP}\%$}: black symbols are the experimental data from Coulter Counter measurements (same in Fig.~\ref{fig:histograms}), blue curve corresponds to the full range of measurements, while the orange curve corresponds to the dynamic range of measurements.
}
\label{fig:histmeasurements}
\end{figure*}
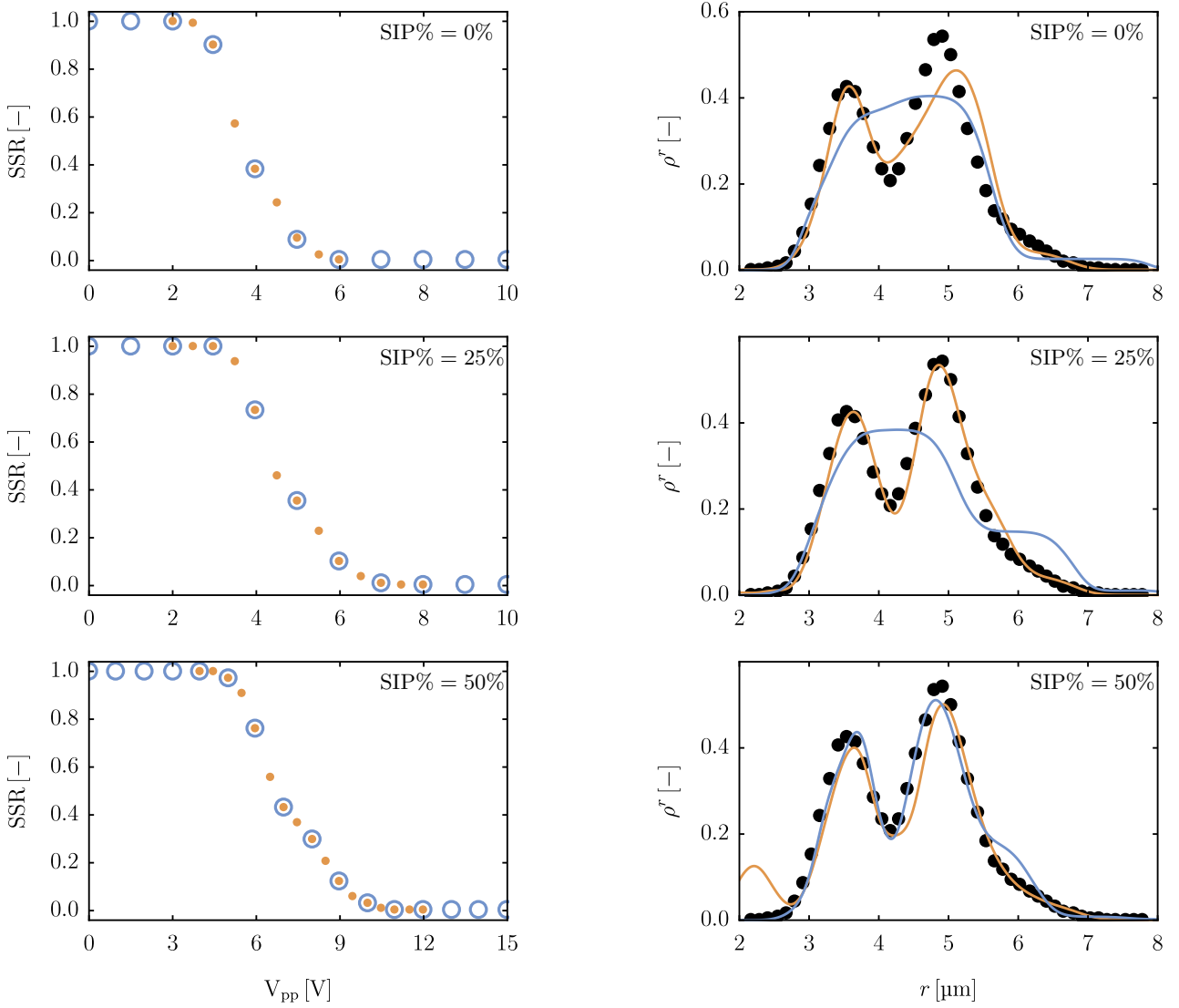

\subsection{Inferring Size Histograms}
Let us suppose one wants to determine the weights for the gaussians that span the parameter space by performing separation experiments at different voltages.
The theoretical value for the side-stream recovery parametrized with respect to the unknown weights $\bm{w^p_{}}$ can be written as
\begin{equation}
\mathrm{SSR}(V;\,\bm{w^p_{}})=\sum_{k\in\mathcal{K}}w^{\bm p}_k\,\mathrm{SSR}_k^{}(V)\,,
\end{equation}
where $V$ is the voltage. When a set of measurements $\mathrm{SSR^{exp}_l}$ as function of the voltage applied on the transducer $V_l$ is available, the set of unknown weights can be determine by requesting that the distance between the measurements and the values provided by MCDGM is minimum
\begin{equation}\label{eq:lsq}
\sum_{l}\left[\sum_{k\in\mathcal{K}}F_{kl}^{}\,w^{\bm p}_k-\mathrm{SSR}^\mathrm{exp}_l\right]^2=\mathrm{min}\,,
\end{equation}
where the matrix $F_{kl}^{}$ is
\begin{equation}
F_{kl}^{}=\mathrm{SSR}_k^{}(V_l^{})\,,
\end{equation}
and it can be viewed as the transfer function for the $k$-th gaussian when a voltage $V_l^{}$ is applied.
The problem Eq.~\eqref{eq:lsq} can be reformulated as rectangular linear equation
\begin{equation}\label{eq:lsqsys}
\sum_{k\in\mathcal{K}}F_{kl}^{}\,w^{\bm p}_k\approx\mathrm{SSR}^\mathrm{exp}_l\,,
\end{equation}
and the solution is given by
\begin{equation}
\bm{w^p_{}}=\bm{F}^+_{}\bm{\mathrm{SSR}}^\mathrm{exp}_{}\,,
\end{equation}
where \mbox{$\bm{F}^+_{}=(\bm{F}^T_{}\bm{F})^{-1}_{}\bm{F}^T_{}$} is the Moore-Penrose pseudodinverse. Alternatively, the Matlab routine \verb|lsqlin| or the Mathematica routine \verb|LeastSquares| perform the same calculations by providing the matrix $\bm{F}$ and the measurements $\bm{\mathrm{SSR}}^\mathrm{exp}_{}$.

Figure~\ref{fig:histmeasurements} shows the results obtained for the size histogram estimation of WBCs from synthetic SSR experiments generated by particle ensemble simulations at different $\mathrm{SIP}\%$ with $\alpha$ dependent on the $\mathrm{SIP}\%$: \mbox{$\alpha(0\%)\simeq 5.63$}, \mbox{$\alpha(25\%)\simeq 6.25$}, and \mbox{$\alpha(40\%)\simeq 6.87$}. The different concentrations were chosen to enlarge the dynamic voltage range, both the full range and the dynamic range were used to determine the histograms.
The gaussians where chosen so that $\dim\mathcal{K}=25$ in a range $r=2-8\,\si{\micro m}$, namely
\begin{subequations}
\begin{align}
m^r_k&=\,r_\mathrm{min}^{}+k\,\Delta r\,,\qquad k=0...\dim\mathcal{K}\,,\\
\sigma^r_k&=\,\Delta r\,\log 2\,.
\end{align}
\end{subequations}
with \mbox{$r_\mathrm{min}^{}=2\,\si{\micro m}$} and $\Delta r=0.25\,\si{\micro m}$.
As it can see from the figure, with the aid of MCDGM method is possible to estimate the radius distribution from the SSR measurements as function of the voltage. The discrepancies between the histogram used to generate the SSR measurements and that computed from the MCDGM method applied to the dynamic range are small. Using the full range seems to provide subsampled histograms. An exception seems to occur for \mbox{$\mathrm{SIP}\%=50\%$} where the dynamic range gives an additional ``bump'' for smaller radii.

\section{Discussion}\label{sec:discussion}
The MCDGM method have been applied to a variety of situations. In all of these case the method has proven its reliability and robustsness in terms of varying the simulation parameters, providing a good approximation of the spatial marginals, the prediction of the SSR, and in the estimation of the histograms.

With respect to the particle ensemble simulations the MCDGM method can generate results much more faster.
The relative computational costs have been estimated by assuming as reference the 1D-MCDGM equations and resulted to be
$7-10\times$ for the 2D-MCDGM equations, and $10\mathrm{k}-15\mathrm{k}\times$ for the particle ensemble simulations.
The exact cost depend on the number of gaussians used in the MCDGM method or the number of particles in the ensemble simulations, the dimensionality of the problem, i.e. 1D, 2D or 3D. It is however possible to claim that the speedup of the MCDGM method over the particle ensemble simulations is about three order of magnitude.

The advantage of this speedup can seem not beneficial for the cases presented in Sec.~\ref{sec:examples}(A)-(B) for which the analytical solution of the lateral movement can be applied to particle ensembles~\cite{simon2017particle}.
It's however remarkable that when the MCDGM method is applied to the simple models in Sec.~\ref{sec:examples}(A)-(B), analytical solutions similar to that presented in \cite{Garofalo_2014} for the dispersion problem are available.
Therefore, the analysis of the computational speedup should been performed based on the analytical solutions, but the results of the comparison are trivial since the MCDGM method can capture the behavior of the PDF with just a few gaussians.
The advantage of using the MCDGM method is however obvious when applied to the models investigated in Sec.~\ref{sec:examples}(C)-(D),
especially in the case of histogram estimations where a significant number of gaussians must be used to have a fine sampling of the parameter space.
%\cite{spencer2014microfluidic}
%\cite{holmes2009leukocytes}

\section{Conclusions}\label{sec:conclusions}
A method for quantifying the acoustophoretic separation of microparticle populations with continously-distributed parameters has been introduced.
The method has been applied to an one-dimensional abstract model of acoustophoretic separation, where the particles were considerd to have a radius distribution.
The approximation property of the method has been illustrated by comparing the statistics for the particle ensemble simulations and those computed by using the MCDGM method.
The application of the method to a model related with previously published experiments of WBC/RBC separation has shown its robustness with respect to distributions and changes in the fluid properties occurring in real-world applications.
Furthermore, the method has been employed to quantify free-flow acoustophoretic separation performance with and without prefocusing of the particle streams, and for the estimation of size histograms from separation performance data.
In all of the cases here investigated the application of the MCDGM method to the model equations has shown very good results in terms of approximation of the numerical data from particle-ensemble simulations and in the estimation of the size histograms.

For that regards future comparisons with experimental data, the MCDGM method promises undisputed advantages for the experimentalists in terms of analysis of the experimental outcomes.
Firstly, it is model-free, meaning that it is possible to increase the complexity of the physical model to obtain a more refined representation and a better consistency with the experimental data.
The complexity can reach the level of numerically synthesized velocity fields, acoustic fields, and precomputed scattering laws on particles of arbitrary shape, while the applicability of the MCDGM method is still guaranteed.
Secondly, it can be adapted to perform both hydrodynamics and acoustics calibration of acoustophoretic devices, so that the model inputs can be actually measured instead of being derived from approximate theoretical estimates and guesses such as those used in this manuscript.

The introduction of the MCDGM method and its application to free-flow acoustophoresis represents a breakthrough for the assessment of the separation performance in acoustophoretic device.
It possesses unprecedented features such as incorporating and estimating parameter- and spatial-distributions, very low computational cost compared to particle ensemble simulations, and effective/practical dimensional reduction.
This means that the numerical implementations of this method are suitable to be executed on single-board computers, enabling thus for ultra-compact applications which embed control, calibration, and analysis algorithms on the same processing unit.

\begin{acknowledgments}
This work was supported by the Knut and Alice Wallenberg
Foundation (Grant No. KAW 2012.0023).
%The author would like to thank Prof.~Thomas Laurell, Andreas Lenshof, Per Augustsson, Pelle Ohlsson, Kevin Cushing, Anke Urbansky and Franziska Olm for the useful discussions about the 
\end{acknowledgments}

\bibliography{acoustofluidics2}

\end{document}